\documentclass{aa}  
\makeatletter
\renewcommand*\aa@pageof{, page \thepage{} of \pageref*{LastPage}}
\usepackage{natbib}
\bibpunct{(}{)}{;}{a}{}{,} 

\usepackage{enumitem}

\usepackage{graphicx}
\usepackage{txfonts}
\usepackage{booktabs}
\usepackage{lipsum}
\usepackage[colorlinks=true,allcolors=blue]{hyperref}
\usepackage{booktabs}
\usepackage{multirow}

\begin{document}

   \titlerunning{Mitigating incoherent excess variance in high-redshift 21-cm observations with cross-GPR}
   \title{Mitigating incoherent excess variance in high-redshift 21 cm observations with multi-output cross-Gaussian process regression}

   \subtitle{}

   \author{S. Munshi\inst{1}
          \and
          L. V. E. Koopmans\inst{1}
          \and
          F. G. Mertens\inst{2,1}
          \and
          A. R. Offringa\inst{3,1}
          \and 
          S. A. Brackenhoff\inst{1} 
          \and 
          E. Ceccotti\inst{4,1} 
          \and 
          J. K. Chege\inst{1}
          \and 
          L. Y. Gao\inst{5,1} 
          \and 
          S. Ghosh\inst{1}
          \and
          M. Mevius\inst{3}
          \and
          S. Zaroubi\inst{1,6}
          }

   \institute{Kapteyn Astronomical Institute, University of Groningen, P.O. Box 800, 9700 AV Groningen, The Netherlands\\ \email{munshi@astro.rug.nl}
         \and
    LUX, Observatoire de Paris, Universit\'{e} PSL, CNRS, Sorbonne Universit\'{e}, F-75014 Paris, France
        \and
    ASTRON, PO Box 2, 7990 AA Dwingeloo, The Netherlands
        \and 
    INAF -- Istituto di Radioastronomia, Via P.~Gobetti 101, 40129 Bologna, Italy
        \and
    School of Science, Jiangxi University of Science and Technology, Ganzhou 341000, China
        \and
    ARCO (Astrophysics Research Center), Department of Natural Sciences, The Open University of Israel, 1 University Road, PO Box 808, Ra’anana 4353701, Israel
    }

   \date{Received --; accepted --}
 
  \abstract
    {Systematic effects that limit the achievable sensitivity of current low-frequency radio telescopes to the 21 cm signal are among the foremost challenges in observational 21 cm cosmology. The standard approach to retrieving the 21 cm signal from radio interferometric data separates it from bright astrophysical foregrounds by exploiting their spectrally smooth nature, in contrast to the finer spectral structure of the 21 cm signal. Contaminants exhibiting rapid frequency fluctuations, on the other hand, are difficult to separate from the 21 cm signal using standard techniques and the power from these contaminants contributes to low-level systematics that can limit our ability to detect the 21 cm signal. Many of these low-level systematics are incoherent across multiple nights of observation, resulting in an incoherent excess variance above the thermal noise sensitivity of the instrument. In this work, we developed a method called  cross-covariance Gaussian process regression (cross-GPR) that exploits the incoherence of these systematics to separate them from the 21 cm signal, which remains coherent across multiple nights of observation. We  developed and demonstrated the technique on synthetic signals in a general setting,  then we applied it to gridded interferometric visibility cubes. We performed realistic simulations of visibility cubes containing foregrounds, 21 cm signal, noise, and incoherent systematics. The simulations show that the method can successfully separate and subtract incoherent contributions to the excess variance. Furthermore, its advantages over standard techniques become more evident when the spectral behavior of the contaminants resembles that of the 21 cm signal. Simulations performed on a variety of 21 cm signal shapes also reveal that the cross-GPR approach can subtract incoherent contributions to the excess variance, without suppressing the 21 cm signal. The codes underlying this article are publicly available in the Python library \texttt{crossgp} and will soon be integrated into the LOFAR and NenuFAR foreground removal and power spectrum estimation framework \texttt{ps\_eor}.}


   \maketitle
%

\section{Introduction}\label{sec:introduction}
One of the most promising direct probes of the high-redshift Universe is the 21 cm signal from neutral hydrogen. The three-dimensional spatial fluctuations of the 21 cm brightness temperature at high redshifts can be probed by low-frequency radio interferometers. Several instruments, such as  GMRT\footnote{Giant Metrewave Radio Telescope} \citep{paciga2013simulation}, LOFAR\footnote{Low-Frequency Array} \citep{patil2017upper,gehlot2019first,gehlot2020aartfaac,mertens2020improved,mertens2025deeper,ceccotti2025first}, MWA\footnote{Murchison Widefield Array}  \citep{ewall2016first,barry2019improving,li2019first,trott2020deep,yoshiura2021new,nunhokee2025limits}, PAPER\footnote{Precision Array to Probe EoR} \citep{kolopanis2019simplified}, HERA\footnote{Hydrogen Epoch of Reionization Array} \citep{abdurashidova2022first,adams2023improved}, OVRO-LWA\footnote{Owens Valley Radio Observatory - Long Wavelength Array} \citep{eastwood201921,garsden202121}, and NenuFAR\footnote{New Extension in Nan\c cay Upgrading LOFAR} \citep{munshi2024first,munshi2025improved} have attempted (or are currently attempting) to detect the 21 cm signal from  cosmic dawn and the epoch of reionisation, two critical periods in the high-redshift Universe. The collecting area of the current generation of experiments only allows them to probe the 21 cm signal statistically in reasonable observing hours, by measuring the signal power spectrum. However, due to the steep technical challenges of recovering the faint signal from below orders of magnitude brighter foregrounds, none of these instruments have yet detected it; however, they have been able to set increasingly stringent upper limits on the 21 cm signal power spectrum.

Even if the thermal noise has been reduced to below the 21 cm signal level, there are other obstacles that prevent us from utilising the full sensitivity of the instrument necessary for detecting the 21 cm signal fluctuations. The foremost amongst them are the astrophysical foregrounds, due to Galactic and extragalactic emission. Low-frequency radio foregrounds are several orders of magnitude brighter than the expected background 21 cm signal. This creates a very steep calibration challenge, imposing stringent precision requirements on the calibration algorithms. The difficulty is amplified by the fact that there is usually poor prior knowledge of the instrumental primary beam, particularly near nulls and sidelobes far from the target field \citep{chokshi2024necessity,brackenhoff2025robust}. The residual power from off-axis sources has been identified as one of the main causes of the excess variance above the thermal noise experienced by phase-tracking instruments such as LOFAR, NenuFAR, and MWA. For HERA, on the other hand, cross-coupling systematics have been identified as the primary limiting factor \citep{kern2019mitigating,kern2020mitigating}. There are multiple additional contaminants, such as ionospheric distortions \citep{vedantham2015scintillation,vedantham2016scintillation,brackenhoff2024ionospheric}, incomplete sky models \citep{barry2016calibration,hofer2025impact}, inaccurate foreground source modeling \citep{ceccotti2025spectral}, and low-level radio frequency interference \citep{offringa2019impact,wilensky2019absolving}, which could pose systematic limitations to our ability to utilise the full sensitivity of the instruments.

The biggest challenge in observational 21 cm cosmology, however, is foreground mitigation. Broadly, foreground mitigation strategies fall into two categories: avoidance and subtraction. Foreground avoidance techniques exclude the foreground wedge, a region in the $(k_{\perp}, k_{\parallel})$ space where foregrounds are predominantly confined \citep{datta2010bright,vedantham2012imaging,morales2012four,munshi2025beyond}, while estimating the power spectrum of the 21 cm signal. Foreground subtraction techniques, on the other hand, use the spectral smoothness of foregrounds as priors in signal separation algorithms to model and remove foreground contamination from the data. Several foreground subtraction approaches, both parametric and non-parametric, have been proposed, among which Gaussian process regression \citep[GPR;][]{mertens2018statistical,mertens2024retrieving} has emerged as one of the most effective methods \citep{10.1093/mnras/staf1466}. GPR models the visibility data as the sum of Gaussian processes (GPs) representing foregrounds, the 21 cm signal, noise, and any other component in the data. The spectral coherence of each component is encoded through its respective frequency-frequency covariance function, and the best-fit foregrounds are subtracted from the data.

The presently explored approaches to signal separation in 21 cm cosmology rely on the relative spectral smoothness of foregrounds, compared to the 21 cm signal, and do not attempt to subtract contaminants with rapid spectral fluctuations. However, several contaminating effects may not exhibit a smooth spectral signature, such as terrestrial or satellite radio frequency interference \citep{munshi2025near,di2023unintended,grigg2023detection,gehlot2024transient}, the turbulent ionosphere causing calibration errors on longer baselines to leak into shorter baselines \citep{brackenhoff2024ionospheric}, and calibration errors arising from primary beam inaccuracies near nulls \citep{gan2022statistical,gan2023assessing,brackenhoff2025robust}, all of which can contribute to excess variance in the data. Foreground mitigation algorithms based on spectral smoothness cannot easily separate these contributions from the 21 cm signal, leading to excess variance in the data that limits instrumental sensitivity. Many of these contaminants are incoherent across nights due to their stochastic nature, so their contribution averages out, with the resulting variance decreasing as more nights are combined. The excess variance observed by instruments such as LOFAR and NenuFAR has indeed been seen to integrate down over time \citep{mertens2020improved,mertens2025deeper,munshi2025improved}, suggesting that a significant portion is incoherent, manifesting as excess variance above the theoretical thermal sensitivity of the instrument. In contrast, the observed 21 cm signal within the main lobe of the primary beam remains coherent across multiple nights. This contrast presents an opportunity: the night-to-night incoherence can be used as a property that can help the separation of the contributions of these contaminants from the 21 cm signal contributions to the visibility data.

In many applications of GPR, we encounter multiple correlated outputs, such as repeated measurements and related signal components, which has motivated the development of multi-output GPs that aim to jointly model such outputs, incorporating information on how the outputs covary \citep[e.g.,][]{alvarez2012kernels}. The development of such models was largely driven by the field of geostatistics, where they are known as co-kriging methods. A central goal of these frameworks is to model the cross-covariance, which encodes how different groups of signals are correlated across outputs. Several frameworks have been proposed to model correlations between multiple outputs by expressing each output as a combination of shared latent processes, with the goal of capturing the cross-covariance structure between outputs. For example, the intrinsic co-regionalisation model \citep{goovaerts1997geostatistics} is the simplest, assuming that all outputs are scaled versions of the same latent function. The semiparametric latent factor model \citep[SLFM:][]{teh2005semiparametric} extends this by allowing each output to depend on a linear combination of multiple shared latent functions, each modeled as a GP with a distinct covariance. The linear model of co-regionalisation \citep[LMC:][]{journel1976mining,goovaerts1997geostatistics} generalises both by allowing the outputs to depend not only on multiple GP covariances, but also on multiple independent samples from each. Describing such cross-covariance structures between the outputs of GPs is well-suited to modeling visibility data from multiple nights of radio interferometric observations, where signal components may exhibit coherent or incoherent behavior across nights.

In this paper, we extend the frequency-frequency-only GPR method introduced in \cite{mertens2018statistical} to also account for temporal night-to-night (in)coherence. The extended model captures not only the covariance within an individual dataset, but also the cross-covariance between pairs of datasets, enabling it to capture the (in)coherence of specific components across multiple datasets. We apply this method to signal separation in 21 cm cosmology, using the temporal coherence of the 21 cm signal across multiple nights of observation to separate it from time-incoherent contributions to the excess variance. Subtracting such incoherent contributions along with the foreground component could bring us much closer to the instrumental thermal noise sensitivity and reduce the risk of suppressing the 21 cm signal, which is coherent between nights. Additionally, compared to the analysis by \cite{acharya2024revised} where a bias correction was performed for the excess, such a subtraction not only reduces the bias introduced by the excess variance but also decreases its sample variance, which often dominates the upper limits in 21 cm analyses.

The paper is organised as follows. Section \ref{sec:background} gives an overview of GPR and its application to signal separation. In Sect.~\ref{sec:joint_gpr_multiple}, we describe the new method of separating signals based on coherence in a general setting and then set it in the context of 21 cm signal extraction. Section \ref{sec:simulations} demonstrates the method on simulated radio interferometric visibility cubes. In Sect.~\ref{sec:discussion}, we present the main results, and discuss the applicability and limitations of the current implementation of the method. Section \ref{sec:conclusion} summarises the main conclusions from this paper. We summarise the main mathematical notations used in this paper in Table \ref{tab:notation}.

\section{Gaussian process regression}\label{sec:background}
Gaussian processes (GPs) are widely used tools for modeling signals in noisy data.  They are particularly useful in situations where the underlying functional form of the signal is unknown, but statistical properties such as spectral, temporal, or spatial smoothness can be described. A GP is a probability distribution over functions that can model a set of data points such that the values of a function $f$ evaluated at any finite collection of inputs follow a multivariate normal distribution \citep{Rasmussen2006Gaussian}. Specifically, the function values $\mathbf{f} = f(\mathbf{x})$, evaluated at a vector of inputs $\mathbf{x}$, are distributed as
\begin{equation}
\mathbf{f} \sim \mathcal{GP}(m(\mathbf{x}), \mathbf{K}) = \mathcal{N}(m(\mathbf{x}), \mathbf{K}(\mathbf{x}, \mathbf{x})),
\end{equation}
where $m(\mathbf{x})$ is the mean function, defined component-wise as $m(x_i) = \mathbb{E}[f(x_i)]$, usually assumed to be zero. $\mathbf{K}(\mathbf{x},\mathbf{x})$ is the covariance matrix with entries $\mathbf{K}_{ij}$ given by a positive-definite covariance $\kappa(x_i, x_j)$ between function values at the points $x_i$ and $x_j$.

\begin{table}[t]
\centering
\caption{Summary of scientific notation.}
\begin{tabular}{ll}
\toprule
Symbol & Meaning \\
\midrule
$\mathcal{N}(\mu, \Sigma)$ & Normal distribution with mean $\mu$, covariance $\Sigma$ \\
$\mathbb{E}[X]$ & Expectation value of $X$ \\
$\operatorname{cov}(X, Y)$ & Covariance between $X$ and $Y$ \\
$\mathbf{x}$ & Column vector of input variable values\\
$\mathbf{f}$ & Column vector of function values at input  \\
$\mathbf{d}$ & Column vector of observed data \\
$\mathbf{n}$ & Column vector of noise \\
$\boldsymbol{\theta}$ & Set of hyperparameters \\
$\mathbf{K}$ & Covariance matrix \\
$\sigma^2$ & Variance of GP covariance \\
$\ell$ & Lengthscale of GP covariance \\
$\otimes$ & Kronecker product \\
$\mathbf{I}_N$ & Identity matrix ($N \times N$) \\
$\mathbf{J}_N$ & All-ones matrix ($N \times N$) \\
$\mathbf{1}_N$ & Column vector of ones ($N \times 1$) \\
$k$ & Radial Fourier mode \\
$k_{\parallel}$ & Line-of-sight Fourier mode \\
$k_{\perp}$ & Transverse Fourier mode \\
$\Delta^2$ & Dimensionless power spectrum \\
\bottomrule
\end{tabular}
\label{tab:notation}
\end{table}

\subsection{GPR for signal separation}\label{sec:background_sep}
In real situations, the observed data $\mathbf{d}$ carries an additive noise $\mathbf{n}$ at each set of inputs, such that $\mathbf{d} = \mathbf{f} + \mathbf{n}$ where $\mathbf{n} \sim \mathcal{N}(0, \sigma_n^2\mathbf{I}_p)$, $\sigma_n^2$ being the noise variance and $\mathbf{I}_p \in \mathbb{R}^{p \times p}$ being an identity matrix\footnote{Instead of a scaled identity matrix, any other noise covariance matrix $\mathbf{K}_\mathrm{noise}$ could also be used.} (where $p$ is the number of elements in $\mathbf{x}$). GPR is particularly well-suited for signal separation tasks, where the observed data are modeled as a sum of multiple components. Following the framework developed by \cite{mertens2018statistical}, the total covariance function can then be expressed as a sum of individual GP covariance functions, each capturing the structure of a distinct signal component. If we consider $M$ components, $\mathbf{f}_i$, with corresponding covariances, $\mathbf{K}_i$, the data vector and the GP model covariance are given by
\begin{align}\label{eq:signal_sep}
\mathbf{d} &= \sum_{i=1}^M\mathbf{f}_i+\mathbf{n},\nonumber\\
\mathbf{K} &= \sum_{i=1}^M \mathbf{K}_i.
\end{align}
The joint probability distribution of the observed data and the predicted values of the $k-$th component is then given by
\begin{equation}
\begin{bmatrix} \mathbf{d} \\ \mathbf{f}_k \end{bmatrix}
\sim \mathcal{N} \left(
\begin{bmatrix}0 \\ 0 \end{bmatrix},
\begin{bmatrix}
\mathbf{K} + \sigma_n^2 \mathbf{I}_p & \mathbf{K}_k \\
\mathbf{K}_k & \mathbf{K}_k
\end{bmatrix}
\right).
\end{equation}
GPR estimates the  predictive distribution for $\mathbf{f}_k$ conditioned on $\mathbf{d}$, which is given by
\begin{equation}
\mathbf{f}_k \mid \mathbf{d}, \mathbf{x} \sim \mathcal{N}\left(\mathbb{E}\left(\mathbf{f}_k\right), \operatorname{cov}\left(\mathbf{f}_k\right)\right).
\end{equation}
The predictive distribution is described by the mean $\mathbb{E}(\mathbf{f}_k)$ and the covariance cov($\mathbf{f}_k$), which have the following expressions:
\begin{align}\label{eq:pred_sep}
\mathbb{E}\left(\mathbf{f}_k\right) &=\mathbf{K}_k\left[\mathbf{K}+\sigma_n^2 \mathbf{I}_p\right]^{-1}\mathbf{d},\nonumber\\
\operatorname{cov}\left(\mathbf{f}_k\right) &=\mathbf{K}_k-\mathbf{K}_k\left[\mathbf{K}+\sigma_n^2 \mathbf{I}_p\right]^{-1} \mathbf{K}_k .
\end{align}
Given we have complete knowledge of the covariance, these equations enable us to predict function values for a specific signal component. 

The covariance kernels must be estimated from the information in the data itself. The standard approach is to define analytical covariance functions, which are described by a set of hyperparameters that control properties such as the variance and the coherence scale. Matern functions \citep{matern1960spatial} form a class of such commonly used covariance functions in GPR and are defined as
\begin{equation}\label{eq:matern}
\kappa_{\mathrm{Matern}}(x_{i},x_{j}) = \sigma^{2}\dfrac{2^{1-\eta}}{\Gamma(\eta)}\left(\dfrac{\sqrt{2\eta}r}{\ell}\right)^{\eta}K_{\eta}\left(\dfrac{\sqrt{2\eta}r}{\ell}\right).
\end{equation}
Here, $\sigma^{2}$ is the variance, $\ell$ is the coherence scale, $r=|x_i-x_j|$ is the separation in $\mathbf{x}$, $\Gamma$ is the Gamma function and $K_{\eta}$ is the modified Bessel function of the second kind. The parameter $\eta$ specifies the functional form, with some commonly used values being $\eta=\infty$: radial basis function (RBF) or Gaussian, $\eta=2.5$: Matern 5/2, $\eta=1.5$: Matern 3/2 and $\eta=0.5$: Exponential. The variance describes the span of the function values, while the coherence scale describes how quickly the correlation between the function values at a pair of points in $\mathbf{x}$ drops as we increase the distance between the points. The full covariance of the data then becomes a parametrised function over the set of hyperparameters $\boldsymbol{\theta}$, and the best-fit covariance function can be estimated from the data itself in a Bayesian framework. The vector $\boldsymbol{\theta}$ is composed of $M$ sets of hyperparameters, each set describing $\mathbf{K}_i$. Incorporating priors on the hyperparameter estimates $p(\boldsymbol{\theta})$, the posterior distribution of the hyperparameters is obtained from Bayes' theorem as
\begin{equation}\label{eq:bayes}
\log p(\boldsymbol{\theta} \mid \mathbf{d}, \mathbf{x}) \propto \log p(\mathbf{d} \mid \mathbf{x}, \boldsymbol{\theta})+\log p(\boldsymbol{\theta}).
\end{equation}
Here $\log p(\mathbf{d}\mid\mathbf{x},\boldsymbol{\theta})$ is the log-marginal-likelihood (LML) function. For a Gaussian likelihood, the LML can be calculated explicitly and is given by
\begin{align}
\log p(\mathbf{d} \mid \mathbf{x}, \boldsymbol{\theta})= -\frac{1}{2} \mathbf{d}^{\top}\left(\mathbf{K}+\sigma_{\mathrm{n}}^2 \mathbf{I}_p\right)^{-1} \mathbf{d}&-\frac{1}{2} \log \left|\mathbf{K}+\sigma_{\mathrm{n}}^2 \mathbf{I}_p\right|\nonumber\\
&-\frac{p}{2} \log 2 \pi.
\end{align}
The covariance model selection is thus performed in a Bayesian framework, by maximising the LML. The posterior distribution of the hyperparameters can be sampled using a Markov chain Monte Carlo (MCMC) routine. The covariance kernels evaluated at a sample $\boldsymbol{\theta}$ in the posterior distribution can now be inserted into Eq.~(\ref{eq:pred_sep}) and the function values for the $k-$th component can be sampled from this predictive distribution, thus allowing us to separate out the $k$-th signal component and determine its full covariance matrix.

\subsection{GPR for foreground subtraction in 21 cm cosmology}\label{sec:standard_approach}
 In the context of 21 cm cosmology, GPR utilises the spectral smoothness of foregrounds to separate them from the 21 cm signal, which has a small frequency coherence scale. In this section, we describe this approach to subtracting foregrounds, developed by \cite{mertens2018statistical}, which has been applied to LOFAR data by \cite{gehlot2019first,gehlot2020aartfaac,mertens2020improved,mertens2025deeper,ceccotti2025first} and NenuFAR data by \cite{munshi2024first,munshi2025improved}. The visibility data are modeled as a sum of functions describing the foregrounds ($\mathbf{f}_{\mathrm{fg}}$), 21 cm signal ($\mathbf{f}_{21}$), and noise ($\mathbf{n}$). Often, it is insufficient to describe the data with just these components, and an additional excess variance component ($\mathbf{f}_{\mathrm{ex}}$) is needed to account for this residual variance from systematic effects. GPR is applied to the gridded complex visibility cube in the $u\varv\nu$ space, along the frequency direction. Thus, in this specific case, $\mathbf{x}$ corresponds to the frequency channels. All real and imaginary visibility components across $u\varv$ cells are assumed to follow the same GPR model, while the kernel hyperparameters are inferred jointly from the full dataset. This modeling was further extended by \cite{mertens2024retrieving}, where the covariance was explicitly formulated as a function of baseline length. In this framework, the visibility data $\mathbf{d}$ can be written as
\begin{equation}\label{eq:standard_comps}
    \mathbf{d} = \mathbf{f}_{\mathrm{fg}}+\mathbf{f}_{21}+\mathbf{f}_{\mathrm{ex}}+\mathbf{n},
\end{equation}
Assuming the components to be uncorrelated, the covariance of the GP model ($\mathbf{K}$) is then the sum of the covariances of the components
\begin{equation}\label{eq:standard_cov}
    \mathbf{K} = \mathbf{K}_{\mathrm{fg}}+\mathbf{K}_{21}+\mathbf{K}_{\mathrm{ex}},
\end{equation}
where $\mathbf{K}_{\mathrm{fg}}$, $\mathbf{K}_{21}$, and $\mathbf{K}_{\mathrm{ex}}$ are the covariances of the foreground, 21 cm signal, and excess components, respectively. The foreground covariance kernel and the priors on its lengthscale are chosen in such a way as to reflect the spectral smoothness of foregrounds. For the 21 cm signal, the covariance kernel should have a corresponding power spectrum that can describe a wide range of 21 cm signal power spectra. Since analytical covariance functions may struggle to capture the full diversity of 21 cm signal statistics, \cite{mertens2024retrieving} and \cite{acharya202421} introduced a learned kernel approach to GPR-based foreground subtraction. In this method, a variational autoencoder (VAE) is trained on simulations to encode 21 cm power spectrum shapes. The resulting latent space defines a low-dimensional, physically informed parametrisation of the 21 cm covariance kernel, whose parameters are then treated as hyperparameters and inferred during the GPR. The excess covariance is used to capture components with rapid spectral fluctuations, which cannot be confidently distinguished from the 21 cm signal component based on the frequency behavior. The excess component is thus not subtracted from the data, resulting in 21 cm signal power spectrum limits exceeding those expected from thermal noise only. We note that using Matern kernels for the foregrounds and excess components assumes that these functions can adequately describe the covariance of the components in the data. Results by the DOTSS-21cm team in the SKA data challenge 3a \citep{10.1093/mnras/staf1466} are a recent demonstration that Matern kernels used in ML-GPR can describe foregrounds reasonably well. \cite{acharya2024revised} and \cite{ceccotti2025first} showed that subtracting the full GP model from the data makes it largely consistent with noise, implying that the GP covariance model can describe the excess variance in actual data reasonably well. Future analyses should ideally replace both foreground and excess kernels with more physics-driven covariances, once the dominant source of the excess variance is identified.

The predictive distribution of any GP component corresponding to a sample in the posterior distribution of $\boldsymbol{\theta}$ can now be estimated. Comparing the function components and the GP covariance model described in Eqs.~(\ref{eq:standard_comps}) and (\ref{eq:standard_cov}) to those in Eq.~(\ref{eq:signal_sep}), the predictive mean and covariance of the foreground component can be obtained using Eq.~(\ref{eq:pred_sep}) as
\begin{align}\label{eq:pred_fg}
\mathbb{E}\left(\mathbf{f}_{\mathrm{fg}}\right) &=\mathbf{K}_{\mathrm{fg}}(\boldsymbol{\theta})\left[\mathbf{K}(\boldsymbol{\theta})+\sigma_n^2 \mathbf{I}_p\right]^{-1}\mathbf{d},\nonumber\\
\operatorname{cov}\left(\mathbf{f}_{\mathrm{fg}}\right) &=\mathbf{K}_{\mathrm{fg}}(\boldsymbol{\theta})-\mathbf{K}_{\mathrm{fg}}(\boldsymbol{\theta})\left[\mathbf{K}(\boldsymbol{\theta})+\sigma_n^2 \mathbf{I}_p\right]^{-1} \mathbf{K}_{\mathrm{fg}}(\boldsymbol{\theta}) .
\end{align}
Foreground realisations generated from this predictive distribution can now be subtracted from the data cube to obtain foreground-subtracted residual cubes:
\begin{equation}
\mathbf{r}^j = \mathbf{d} - \mathbf{f}^j_\mathrm{fg}(\boldsymbol{\theta}),
\end{equation}
where $\mathbf{r}^j$ and $\mathbf{f}^j_\mathrm{fg}$ are the $j-$th residual cube and foreground realisation, respectively. The foreground subtracted power spectrum and uncertainties can be estimated from an ensemble of such residual data cubes, corresponding to a set of $\boldsymbol{\theta}$ sampled from its posterior distribution. The power spectrum uncertainties estimated from this ensemble of residual cubes will then account for the spread of the posterior distribution in the hyperparameter space, as well as the measurement errors.

In this GP model, among the various components of the visibility data in Eq.~(\ref{eq:standard_comps}), the foregrounds and the 21 cm signal are expected to be coherent across visibility cubes from multiple nights, while the noise is incoherent. The excess component, however, may exhibit both coherent and incoherent contributions across datasets. A GPR framework that operates on multiple datasets can explicitly incorporate this distinction to disentangle the coherent and incoherent contributions to the data. We develop such a framework in the next section.

\section{GPR for multiple datasets}\label{sec:joint_gpr_multiple}
We first describe the framework for performing GPR on multiple datasets in a general setting. We consider two uncorrelated datasets $\mathbf{d}_1$ and $\mathbf{d}_2$ given by $\mathbf{d}_1 = \mathbf{f}_1 + \mathbf{n}_1$ and $\mathbf{d}_2 = \mathbf{f}_2 + \mathbf{n}_2$, with $\mathbf{f}_1$ and $\mathbf{f}_2$ being the corresponding function values and $\mathbf{n}_1$ and $\mathbf{n_2}$ being the noise realisations. We assume that both functions are described by the same covariance function, $\mathbf{K}$, but with different sets of hyperparameters, $\boldsymbol{\theta}_1$ and $\boldsymbol{\theta}_2$. The joint GP model for the datasets is then given by
\begin{equation}\label{eq:joint_ind}
\begin{bmatrix}\mathbf{d}_1 \\\mathbf{d}_2\end{bmatrix} \sim \mathcal{N}\left(\begin{bmatrix}0 \\0\end{bmatrix},\begin{bmatrix}\mathbf{K}(\boldsymbol{\theta}_1)+\sigma_1^2 \mathbf{I}_p & 0 \\0 & \mathbf{K}(\boldsymbol{\theta}_2)+\sigma_2^2 \mathbf{I}_p\end{bmatrix}\right),
\end{equation}
where $\sigma^2_1$ and $\sigma^2_2$ are the noise variances of the two datasets. This model is equivalent to having two separate GP models for the two datasets, and the distributions of $\boldsymbol{\theta}_1$ and $\boldsymbol{\theta}_2$ can be obtained in the same manner as described in Sect.~\ref{sec:background_sep}, either by performing two separate GPR runs, or by performing a single GPR run with the data and covariances given by Eq.~(\ref{eq:joint_ind}).

\subsection{Separating coherent and incoherent components with cross-GPR}\label{sec:coh_inc_separate}
Given the framework of Eq.~(\ref{eq:joint_ind}), we can now make one further assumption, namely, that the hyperparameters describing the two datasets are also the same, that is, $\boldsymbol{\theta}_1=\boldsymbol{\theta}_2=\boldsymbol{\theta}$. This implies that both datasets can be described by the same covariance kernel.
\begin{equation}\label{eq:joint_link}
\begin{bmatrix}\mathbf{d}_1 \\\mathbf{d}_2\end{bmatrix} \sim \mathcal{N}\left(\begin{bmatrix}0 \\0\end{bmatrix},\begin{bmatrix}\mathbf{K}(\boldsymbol{\theta})+\sigma_1^2 \mathbf{I}_p & 0 \\0 & \mathbf{K}(\boldsymbol{\theta})+\sigma_2^2 \mathbf{I}_p\end{bmatrix}\right),
\end{equation}
Performing a GPR with this model now links the hyperparameters of the two datasets together, and gives a single posterior distribution of $\boldsymbol{\theta}$. However, since the off-diagonal blocks in the covariance matrix are zero, the correlation of the component realisations between the two datasets is still not modeled. The predictive mean estimated from Eq.~(\ref{eq:pred_sep}) will still, in general, be different for the two datasets, since the data vector, $\mathbf{d}$, now has the two different data column vectors, $\mathbf{d}_1$ and $\mathbf{d}_2$, stacked along the rows.

We next consider a situation where the data are a combination of a coherent and an incoherent component. The coherent component has the same realisation across the two datasets, while the incoherent component has different realisations across the two datasets. Defining $\mathbf{f}_{\mathrm{coh}}$ as the coherent component and $\mathbf{f}_{\mathrm{inc}}$ as the incoherent component, the function values $\mathbf{f}_1$ and $\mathbf{f}_2$ corresponding to the two datasets can be written as
\begin{align}\label{eq:joint_funcs}
&\mathbf{f}_1 = \mathbf{f}_{\mathrm{coh}} + \mathbf{f}_{\mathrm{inc}}^1,\nonumber\\
&\mathbf{f}_2 = \mathbf{f}_{\mathrm{coh}} + \mathbf{f}_{\mathrm{inc}}^2,
\end{align}
where $\mathbf{f}_{\mathrm{inc}}^1$ and $\mathbf{f}_{\mathrm{inc}}^2$ are the realisations of the incoherent component for the two datasets. We note that this is a specific case of the LMC framework described by \cite{alvarez2012kernels}.  We describe this connection explicitly in Appendix \ref{sec:lmc}.

Our aim is to include information that one component is coherent while the other is not, within the joint covariance matrix, to improve the separation of the two components. We continue with the assumption that the hyperparameters of both the coherent component ($\boldsymbol{\theta}_\mathrm{coh}$) and the incoherent component ($\boldsymbol{\theta}_\mathrm{inc}$) are linked between the two datasets. We let $\mathbf{K}_\mathrm{coh}$ and $\mathbf{K}_\mathrm{inc}$ be the covariances of the coherent and incoherent components, respectively. The shared covariance function can be written as
\begin{equation}
\mathbf{K} = \mathbf{K}_\text{coh}(\boldsymbol{\theta}_\text{coh})+\mathbf{K}_\text{inc}(\boldsymbol{\theta}_\text{inc}).
\end{equation}
This goes into the diagonal blocks of the covariance in Eq.~(\ref{eq:joint_link}). However, as long as the off-diagonal blocks are zero, the information about the coherence of one component and the incoherence of another is not included in the covariance structure, and the two datasets are solved for independently. We next include this information by estimating the cross-covariances between the two datasets. The cross-covariance is then given by
\begin{equation}
\begin{aligned}
\mathrm{cov}(\mathbf{f}_1,\mathbf{f}_2) 
&= \mathbb{E} \Big[ \big(\mathbf{f}_\text{coh} + \mathbf{f}_\text{inc}^1\big)\cdot \big(\mathbf{f}_\text{coh} + \mathbf{f}_\text{inc}^2\big) \Big] \\
&\hspace{2cm} - \mathbb{E} \big[\mathbf{f}_\text{coh} + \mathbf{f}_\text{inc}^1\big] \cdot \mathbb{E} \big[\mathbf{f}_\text{coh} + \mathbf{f}_\text{inc}^2\big] \\
&= \mathbf{K}_{\mathrm{coh}}, \ \text{since } \mathbb{E}[\mathbf{f}_{\mathrm{inc}}^1\cdot\mathbf{f}_{\mathrm{inc}}^2] = 0.
\end{aligned}
\end{equation}
Equation~(\ref{eq:joint_link}) thus gets modified to
\begin{align}\label{eq:joint_cross}
\begin{bmatrix}\mathbf{d}_1 \\\mathbf{d}_2\end{bmatrix} &\sim \mathcal{N}\left(\begin{bmatrix}0 \\0\end{bmatrix},\begin{bmatrix}\mathbf{K}_{\mathrm{coh}}+\mathbf{K}_{\mathrm{inc}}+\sigma_1^2 \mathbf{I}_p & \mathbf{K}_{\mathrm{coh}} \\\mathbf{K}_{\mathrm{coh}} & \mathbf{K}_{\mathrm{coh}}+\mathbf{K}_{\mathrm{inc}}+\sigma_2^2 \mathbf{I}_p\end{bmatrix}\right)\nonumber\\
\equiv \mathbf{d} &\sim \mathcal{N}(0, \mathbf{K}_{\mathrm{full}}).
\end{align}
Thus, the GP model now correctly includes the information that one of the components of the model is coherent between the two datasets while the other is incoherent.

The predictive distribution of the coherent and incoherent components corresponding to a sample in the posterior distribution of $\boldsymbol{\theta}=[\boldsymbol{\theta}_\mathrm{coh},\boldsymbol{\theta}_\mathrm{inc}]$ is then given by
\begin{subequations}
\begin{align}
\mathbb{E}(\mathbf{f}_{\mathrm{coh}}) &= \left( \mathbf{1}_2^\top \otimes \mathbf{K}_{\mathrm{coh}}(\boldsymbol{\theta}_\mathrm{coh}) \right) 
\mathbf{K}_{\mathrm{full}}(\boldsymbol{\theta})^{-1} \, \mathbf{d}, \nonumber\\
\mathrm{cov}(\mathbf{f}_{\mathrm{coh}}) &= \mathbf{K}_{\mathrm{coh}}(\boldsymbol{\theta}_\mathrm{coh}) 
- \left( \mathbf{1}_2^\top \otimes \mathbf{K}_{\mathrm{coh}}(\boldsymbol{\theta}_\mathrm{coh}) \right) 
\mathbf{K}_{\mathrm{full}}(\boldsymbol{\theta})^{-1} \nonumber\\
&\hspace{1.8cm} \times \left( \mathbf{1}_2 \otimes \mathbf{K}_{\mathrm{coh}}(\boldsymbol{\theta}_\mathrm{coh}) \right); \label{eq:pred_coh} \\
\mathbb{E}\left(\begin{bmatrix} \mathbf{f}_{\mathrm{inc}}^{1} \\[0.3em] \mathbf{f}_{\mathrm{inc}}^{2} \end{bmatrix}\right) 
&= \left( \mathbf{I}_2 \otimes \mathbf{K}_{\mathrm{inc}}(\boldsymbol{\theta}_\mathrm{inc}) \right) 
\mathbf{K}_{\mathrm{full}}(\boldsymbol{\theta})^{-1} \mathbf{d}, \nonumber\\
\mathrm{cov}\left(\begin{bmatrix} \mathbf{f}_{\mathrm{inc}}^{1} \\[0.3em] \mathbf{f}_{\mathrm{inc}}^{2} \end{bmatrix}\right) 
&=  \mathbf{I}_2 \otimes \mathbf{K}_{\mathrm{inc}}(\boldsymbol{\theta}_\mathrm{inc}) 
- \left( \mathbf{I}_2 \otimes \mathbf{K}_{\mathrm{inc}}(\boldsymbol{\theta}_\mathrm{inc}) \right) 
\mathbf{K}_{\mathrm{full}}(\boldsymbol{\theta})^{-1} \nonumber\\
&\hspace{2cm}\quad \times \left( \mathbf{I}_2 \otimes \mathbf{K}_{\mathrm{inc}}(\boldsymbol{\theta}_\mathrm{inc}) \right), \label{eq:pred_inc}
\end{align}
\end{subequations}
where $\mathbf{1}_2 \in \mathbb{R}^{2}$ is a two-element column vector of ones and $\mathbf{I}_2 \in \mathbb{R}^{2 \times 2}$ is a two-by-two identity matrix. We thus get a single estimate of the coherent component for both datasets and two separate estimates for the incoherent component.

\begin{figure*}
    \includegraphics[width=1\columnwidth]{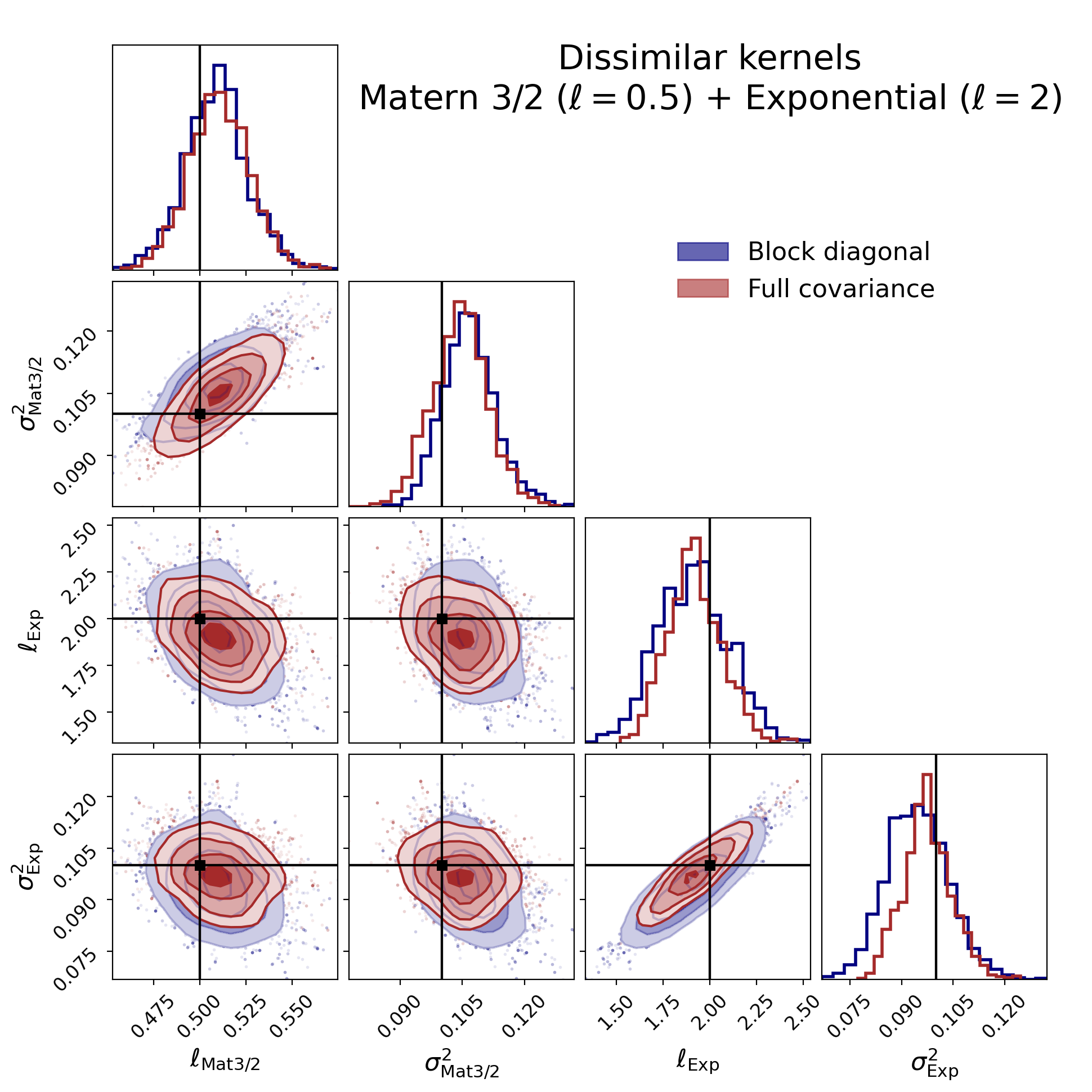}
    \includegraphics[width=1\columnwidth]{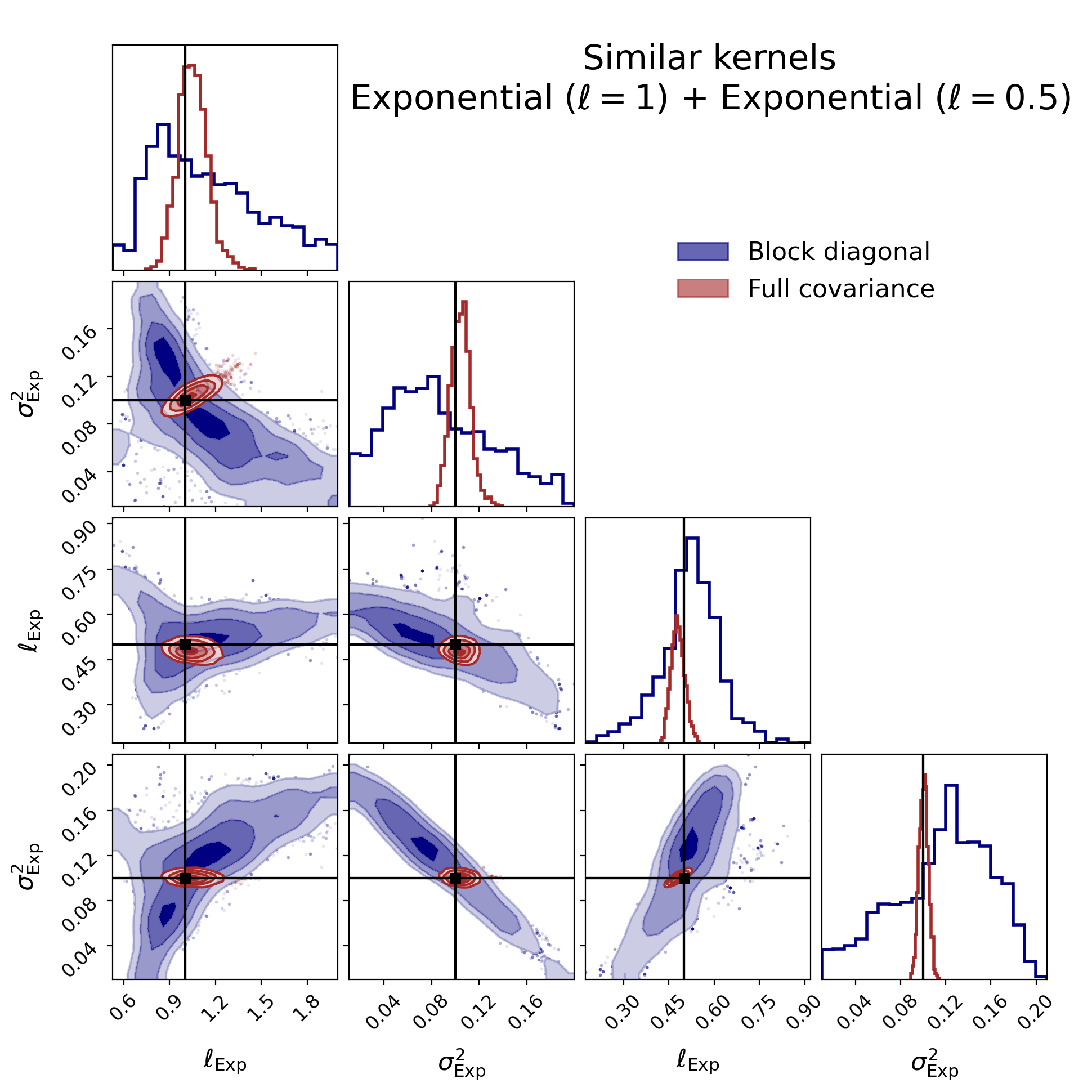}\\
    \par\vspace{0.5cm}
    \includegraphics[width=2\columnwidth]{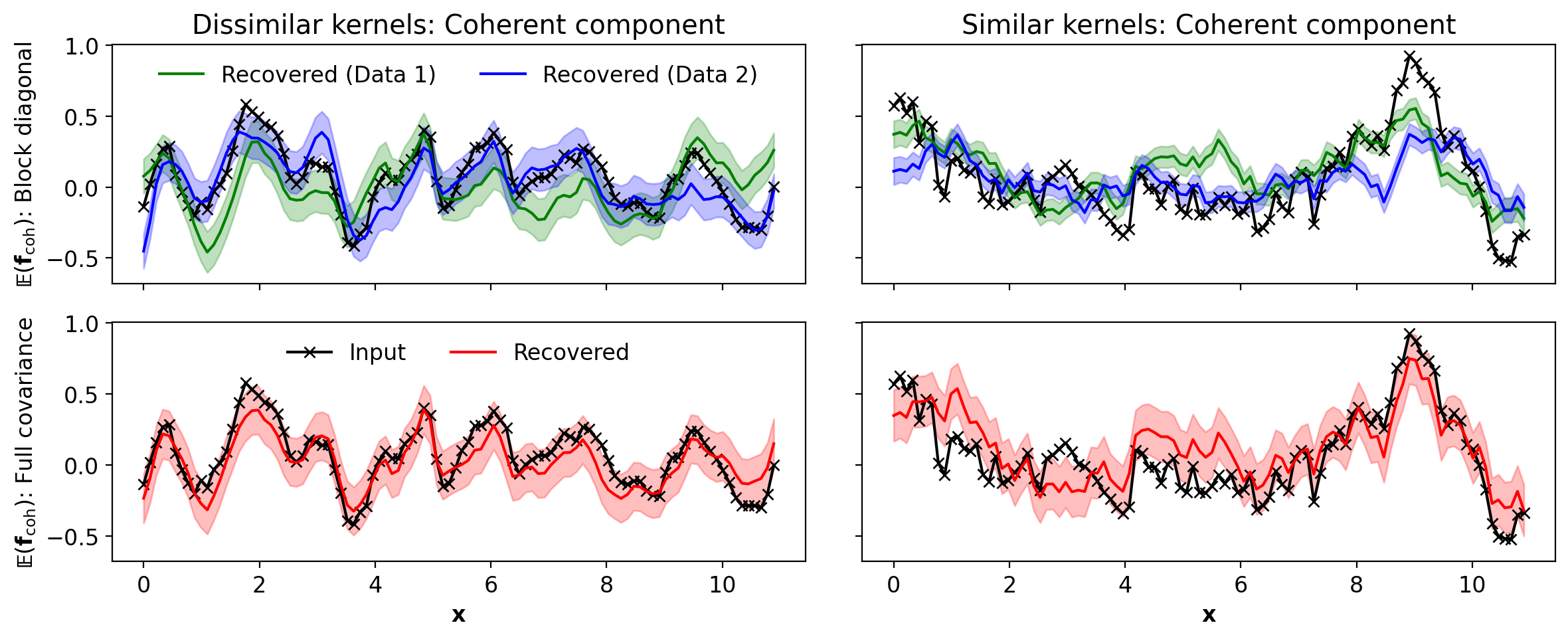}
    \caption{ Impact of including cross-covariances in GPR in the separation of coherent and incoherent components demonstrated on synthetic data. The left column shows the case where the coherent and incoherent covariance kernels are distinct, while the right column shows the case of the coherent and incoherent components having similar covariances. The corner plots in the top show the posterior distribution of the hyperparameters for the block diagonal approach (in blue) and the full covariance approach (in red). The four plots at the bottom show the recovery of a single realisation of the coherent component, for the block diagonal and full covariance approaches, respectively.}
    \label{fig:demo_example}
\end{figure*}
\subsection{Demonstration on synthetic datasets}\label{sec:demo_synthetic}
We next demonstrate how explicitly including the off-diagonal cross-covariance blocks improves our ability to separate out contributions from coherent and incoherent components in the data. For this purpose, we generated two synthetic datasets. Each dataset contains 50 signal realisations, where each signal is composed of two components: a coherent component shared between the two datasets, and an incoherent component that is independent between them. We performed GPR twice, first with the block diagonal GP model of Eq.~(\ref{eq:joint_link}) which assumes the two datasets to be independent, but with linked hyperparameters\footnote{This is equivalent to fitting the average of the two datasets,  a standard approach in the literature to date.}, and second with the full GP model of Eq.~(\ref{eq:joint_cross}) which includes information about the coherence and incoherence of the two components. We sampled the hyperparameter posterior distribution using the MCMC routine \texttt{emcee}\footnote{\url{https://emcee.readthedocs.io/}}, and the hyperparameter estimate with the highest posterior probability was used to calculate the predictive mean and covariance of both the coherent and incoherent components, using Eq.~(\ref{eq:pred_coh}) and Eq.~(\ref{eq:pred_inc}) respectively. We kept the prior range relatively wide and fixed between the two tests. The tests were repeated for a large number of choices of pairs of kernels. We show two such cases:
\begin{enumerate}
    \item Dissimilar kernels: Coherent Matern 3/2 ($\ell=0.5, \sigma^2=0.1$) + Incoherent Exponential ($\ell=2, \sigma^2=0.1$)
    \item Similar kernels: Coherent Exponential ($\ell=1, \sigma^2=0.1$) + Incoherent Exponential ($\ell=0.5, \sigma^2=0.1$)
\end{enumerate}
The results are shown in Fig.~\ref{fig:demo_example}. The dissimilar and similar kernel examples are shown in the left and right columns, respectively. The corner plots in the top show the posterior distribution of the hyperparameters for both block diagonal and full-covariance approaches. For the dissimilar kernels (in the left column), we find that both approaches work relatively well and the posterior distributions recover the input hyperparameter values. We note that the full-covariance posteriors are slightly better constrained. However, for similar kernels (in the right column), we find that the block diagonal approach has wide degenerate posteriors, while the full covariance approach provides much better constrained and less degenerate posteriors around the input hyperparameter values. This is because, in the dissimilar kernels case, the two components are distinct enough to be separable by performing a standalone GPR on each dataset, and including the information that one kernel is coherent while the other is not does not yield significant additional benefits. However, for similar kernels, where the covariance structure along $\mathbf{x}$ is not distinct enough to distinguish the two components, including additional information through the cross-covariance blocks offers significant benefits in recovering the input hyperparameter values. The middle and bottom row shows the predictive mean and standard deviation for a single realisation of the coherent component, obtained at the hyperparameter estimate with the highest posterior. The full covariance approach yields a single coherent component for the two datasets, while the diagonal approach yields one coherent component for each of the two datasets. We find that for the dissimilar kernels, though the full covariance approach can describe the input signal better, the block diagonal approach can also recover the input to some extent. However, for the similar kernels case, the block diagonal approach does significantly worse than the full covariance approach.

\subsection{Generalisation to $N$ datasets}\label{sec:generalization}
The formalism developed in Sect.~\ref{sec:coh_inc_separate} can be naturally extended to a situation where we have a large number of datasets, where the data can be expressed as a sum of coherent and incoherent components. We let $N$ be the number of datasets. Equation~(\ref{eq:joint_cross}) then can be generalised as
\begin{align}\label{eq:joint_cross_general}
\mathbf{d} &\sim \mathcal{N}\left(0,\;
\mathbf{J}_N \otimes \mathbf{K}_{\mathrm{coh}} + \mathbf{I}_N \otimes \mathbf{K}_{\mathrm{inc}} + D \otimes \mathbf{I}_p
\right),
\end{align}
where $\mathbf{J}_N = \mathbf{1}_N \mathbf{1}_N^\top \in \mathbb{R}^{N \times N}$ is an all-ones matrix,  $\mathbf{I}_N \in \mathbb{R}^{N \times N}$ is an identity matrix, and $D = \mathrm{diag}(\sigma_1^2, \dots, \sigma_N^2) \in \mathbb{R}^{N \times N}$ is a diagonal matrix containing the noise variances of individual datasets. For a sample in the posterior distribution of $\boldsymbol{\theta}$, the predictive mean and covariance of the coherent component can be calculated from Eq.~(\ref{eq:pred_coh}), with $\mathbf{1}_2$ replaced by $\mathbf{1}_N$. Similarly, the $N$ incoherent components can be obtained from Eq.~(\ref{eq:pred_inc}), with the column vector of the function values containing $N$ sets of $\mathbf{f}_i$ and $\mathbf{I}_2$ replaced by $\mathbf{I}_N$.

The computational cost of this method, however, increases steeply with the number of datasets $N$. The full covariance matrix $\mathbf{K}_{\mathrm{full}}(\boldsymbol{\theta})$ has dimensions $Np \times Np$. Computing the log marginal likelihood involves a matrix inversion which scales as $\mathcal{O}(N^3 p^3)$. This rapidly becomes a bottleneck as $N$ increases. Since we assume all nights to have the same covariance kernel, a Toeplitz or band-diagonal structure could be employed in the full covariance matrix to speed up the inversion. Alternative formalisms can also include populating the full covariance of $N$ datasets with blocks for pairs of datasets along the diagonal, in which only the (in)coherence between the $2i^{\mathrm{th}}$ and $(2i+1)^{\mathrm{}th}$ datasets is used in performing GPR. However, since we are essentially performing a separate GPR for each pair of datasets, the computational cost reduces to $\mathcal{O}(N p^3)$. Situations where $N>2$ have not been explored in this study and are left for future work.

\subsection{Cross-GPR in 21 cm signal extraction}\label{sec:joint_gpr_21cm}
In this section, we describe the cross-covariance approach in the context of gridded visibility cubes. As described in Sect.~\ref{sec:standard_approach}, in 21 cm cosmology analyses, there are often components in the data in addition to the foregrounds, 21 cm signal, and noise, commonly referred to as excess components. This excess is usually described by a covariance kernel with rapid spectral fluctuations. As a result, the excess component is not subtracted from the data, since it carries a risk of subtracting the 21 cm signal itself. However, in multiple 21 cm cosmology analyses \citep{mertens2020improved,mertens2025deeper,munshi2025improved}, it has been seen that integrating multiple nights of observation not only reduces the thermal noise but also the excess variance. Cross-coherence analyses performed with both LOFAR \citep{mertens2025deeper} and NenuFAR \citep{munshi2025improved} also show that a large portion of the excess component is incoherent across multiple nights of observation, particularly within the EoR window. This incoherent excess variance could have multiple possible origins, such as calibration errors, bright source residuals due to ionospheric effects, and transient radio frequency interference. The 21 cm signal field, on the other hand, is coherent over multiple nights of observations. 

The cross-GPR formalism developed in Sect.~\ref{sec:coh_inc_separate} thus maps ideally onto this problem. The 21 cm signal, foregrounds, and any coherent portion of the excess constitute the coherent components, and the noise and incoherent portion of the excess constitute the incoherent components. The coherent excess could arise from effects such as cross-coupling between stations, persistent RFI, and primary beam model inaccuracies, and is, conservatively, not subtracted from the data. The gridded visibility cubes for two nights of observations $\mathbf{d}_1$ and $\mathbf{d}_2$ are then given by 
\begin{align}
\mathbf{d}_1 &= 
\mathbf{f}_{\mathrm{fg}}+\mathbf{f}_{21}+\mathbf{f}_{\mathrm{ex}^\mathrm{coh}} + \mathbf{f}^{1}_{\mathrm{ex}^\mathrm{inc}}+\mathbf{n}_1,\nonumber\\
\mathbf{d}_2 &= 
\underbrace{\mathbf{f}_{\mathrm{fg}}+\mathbf{f}_{21}+\mathbf{f}_{\mathrm{ex}^\mathrm{coh}}}_{\text{Coherent}} +
\underbrace{\mathbf{f}^{2}_{\mathrm{ex}^\mathrm{inc}}+\mathbf{n}_2.}_{\text{Incoherent}}
\end{align}
Here, $\mathbf{f}_{\mathrm{ex}^\mathrm{coh}}$ is the coherent excess component and $\mathbf{f}^{1}_{\mathrm{ex}^\mathrm{inc}}$ and $\mathbf{f}^{2}_{\mathrm{ex}^\mathrm{inc}}$ are the incoherent excess components from the two nights. Following Eq.~(\ref{eq:joint_cross}), the joint GP model covariance of the two nights is then given by
\begin{equation}
\mathbf{K}_{\mathrm{model}} = \begin{bmatrix}\mathbf{K}_{\mathrm{fg}}+\mathbf{K}_{21}+\mathbf{K}^{\mathrm{coh}}_{\mathrm{ex}}+\mathbf{K}^{\mathrm{inc}}_{\mathrm{ex}}& \mathbf{K}_{\mathrm{fg}}+\mathbf{K}_{21}+\mathbf{K}^{\mathrm{coh}}_{\mathrm{ex}} \\\mathbf{K}_{\mathrm{fg}}+\mathbf{K}_{21}+\mathbf{K}^{\mathrm{coh}}_{\mathrm{ex}} & \mathbf{K}_{\mathrm{fg}}+\mathbf{K}_{21}+\mathbf{K}^{\mathrm{coh}}_{\mathrm{ex}}+\mathbf{K}^{\mathrm{inc}}_{\mathrm{ex}}\end{bmatrix},
\end{equation}
in which the off-diagonal blocks receive contributions only from the coherent components. The joint distribution of the two data cubes is then
\begin{align}
\begin{bmatrix}\mathbf{d}_1 \\\mathbf{d}_2\end{bmatrix} 
&\sim \mathcal{N}\left(
\begin{bmatrix}0 \\0\end{bmatrix},\;
\mathbf{K}_{\mathrm{model}} +
\begin{bmatrix}
\sigma_1^2 \mathbf{I}_p & 0 \\
0 & \sigma_2^2 \mathbf{I}_p
\end{bmatrix}
\right), \nonumber \\
\equiv \mathbf{d} &\sim \mathcal{N}(0, \mathbf{K}_{\mathrm{full}}),
\end{align}
and $\boldsymbol{\theta}$ contains the hyperparameters for all coherent as well as incoherent components. Since the aim is to subtract the foreground component as well as the incoherent excess component from both data cubes, we estimate the predictive distribution of the sum of the foreground and incoherent excess components for the two data cubes. The mean and covariance of this distribution, for the two nights of observations, are given by
\begin{align}\label{eq:pred_fg_inc}
\mathbb{E}\left(
\begin{bmatrix}
\mathbf{f}_{\text{fg}} + \mathbf{f}^{1}_{\mathrm{ex}^{\mathrm{inc}}} \\[0.3em]
\mathbf{f}_{\text{fg}} + \mathbf{f}^{2}_{\mathrm{ex}^{\mathrm{inc}}}
\end{bmatrix}
\right)
 &= \mathbf{K}_{\mathrm{fg},\mathrm{ex}^\mathrm{inc}}(\boldsymbol{\theta}) \mathbf{K}_{\mathrm{full}}(\boldsymbol{\theta})^{-1}\mathrm{d},\nonumber\\
\mathrm{cov}\left(
\begin{bmatrix}
\mathbf{f}_{\text{fg}} + \mathbf{f}^{1}_{\mathrm{ex}^{\mathrm{inc}}} \\[0.3em]
\mathbf{f}_{\text{fg}} + \mathbf{f}^{2}_{\mathrm{ex}^{\mathrm{inc}}}
\end{bmatrix}
\right) &= \mathbf{K}_{\mathrm{fg},\mathrm{ex}^\mathrm{inc}}(\boldsymbol{\theta})-\mathbf{K}_{\mathrm{fg},\mathrm{ex}^\mathrm{inc}}(\boldsymbol{\theta}) \mathbf{K}_{\mathrm{full}}(\boldsymbol{\theta})^{-1}\mathbf{K}_{\mathrm{fg},\mathrm{ex}^\mathrm{inc}}(\boldsymbol{\theta}),\nonumber\\
\text{ where \;} \mathbf{K}_{\mathrm{fg},\mathrm{ex}^\mathrm{inc}}(\boldsymbol{\theta}) &= \begin{bmatrix}
 \mathbf{K}_{\mathrm{fg}}(\boldsymbol{\theta})+\mathbf{K}^{\mathrm{inc}}_{\mathrm{ex}}(\boldsymbol{\theta}) & \mathbf{K}_{\mathrm{fg}}(\boldsymbol{\theta})\\
 \mathbf{K}_{\mathrm{fg}}(\boldsymbol{\theta}) & \mathbf{K}_{\mathrm{fg}}(\boldsymbol{\theta})+\mathbf{K}^{\mathrm{inc}}_{\mathrm{ex}}(\boldsymbol{\theta})
\end{bmatrix}.
\end{align}
This distribution can be used to generate realisations for the sum of foreground and incoherent excess components for both nights. These are subtracted from the input data cubes to yield the foreground and incoherent excess subtracted residual cubes, expressed as
\begin{align}
\mathbf{r}^j_{1} &= \mathbf{d}_1 - \mathbf{f}^j_{\text{fg}}(\boldsymbol{\theta}) - \mathbf{f}^{1,j}_{\mathrm{ex}^{\mathrm{inc}}}(\boldsymbol{\theta}),\nonumber\\
\mathbf{r}^j_{2} &= \mathbf{d}_2 - \mathbf{f}^j_{\text{fg}}(\boldsymbol{\theta}) - \mathbf{f}^{2,j}_{\mathrm{ex}^{\mathrm{inc}}}(\boldsymbol{\theta}),
\end{align}
where $\mathbf{r}^j_{1}$ and $\mathbf{r}^j_{2}$ are the $j-$th residual cubes for the two nights, with $\mathbf{f}^j_{\text{fg}}$ being the coherent foreground component realisation, and $\mathbf{f}^{1,j}_{\mathrm{ex}^{\mathrm{inc}}}$ and $\mathbf{f}^{2,j}_{\mathrm{ex}^{\mathrm{inc}}}$ being the incoherent excess component realisations for the two nights. The power spectrum and uncertainties for either the individual night data cubes or the combined cube are estimated from an ensemble of such residual cubes, produced by sampling the hyperparameter posterior distribution and the predictive distribution of the GP components, as described at the end of Sect.~\ref{sec:standard_approach}. We note that not only the foregrounds and incoherent excess, but any component or sum of components can be recovered similarly, by placing the coherent component covariances in both diagonal and off-diagonal blocks and the incoherent component covariances in only the diagonal blocks.

\begin{table}
\caption{ Covariance model used in GPR with linked hyperparameters for the two nights.}
\label{tab:gpr}
\centering
\renewcommand{\arraystretch}{1.5}
\begin{tabular}{l@{\hskip 4pt}l@{\hskip 4pt}c@{\hskip 4pt}c@{\hskip 4pt}c@{\hskip 4pt}c}
\toprule
\multirow{2}{*}{Kernel} & \multirow{2}{*}{Parm} & \multirow{2}{*}{Input} & \multirow{2}{*}{Prior} & \multicolumn{2}{c}{Posterior} \\
\cmidrule(lr){5-6}
& & & & Standard & Cross \\
\midrule
\multirow{2}{*}{\shortstack[l]{$\mathbf{K}_{\mathrm{int}}$\\(RBF)}} 
  & $\sigma^{2}$ & -0.344 & -1, 0 & $-0.344^{+0.019}_{-0.018}$ & $-0.342^{+0.017}_{-0.019}$ \\
  & $\ell$       & 27.171 & 20, 40 & $26.684^{+0.813}_{-0.751}$ & $26.880^{+0.779}_{-0.815}$ \\
\multirow{2}{*}{\shortstack[l]{$\mathbf{K}_{\mathrm{mix}}$\\(RBF)}} 
  & $\sigma^{2}$ & -2.105 & -2.5, -1.5 & $-2.101^{+0.006}_{-0.006}$ & $-2.104^{+0.005}_{-0.005}$ \\
  & $\ell$       & 0.503  & 0.1, 1 & $0.505^{+0.002}_{-0.002}$ & $0.504^{+0.001}_{-0.001}$ \\
\multirow{3}{*}{\shortstack[l]{$\mathbf{K}_{\mathrm{21}}$\\(VAE)}}  
  & $x_1$        & 1      & -4, 4 & $1.083^{+1.806}_{-2.034}$ & $1.658^{+1.043}_{-0.544}$ \\
  & $x_2$        & -0.5   & -4, 4 & $1.612^{+1.142}_{-0.860}$ & $-0.467^{+0.093}_{-0.053}$ \\
  & $\sigma^{2}$ & -3.449 & -5, -2 & $-4.387^{+0.203}_{-0.128}$ & $-3.510^{+0.039}_{-0.055}$ \\
\multirow{2}{*}{\shortstack[l]{$\mathbf{K}_{\mathrm{ex}}$\\(Exp)}} 
  & $\sigma^{2}$ & -3.960 & -5, -2 & $-3.568^{+0.062}_{-0.088}$ & $-3.959^{+0.004}_{-0.004}$ \\
  & $\ell$       & 0.251  & 0.1, 0.5 & $0.414^{+0.058}_{-0.085}$ & $0.253^{+0.003}_{-0.003}$ \\
\bottomrule
\end{tabular}
\tablefoot{The prior column indicates the uniform prior bounds. The posterior column indicates the estimated median and 68\% confidence intervals obtained with MCMC.}
\end{table}

\begin{figure}
    \includegraphics[width=\columnwidth]{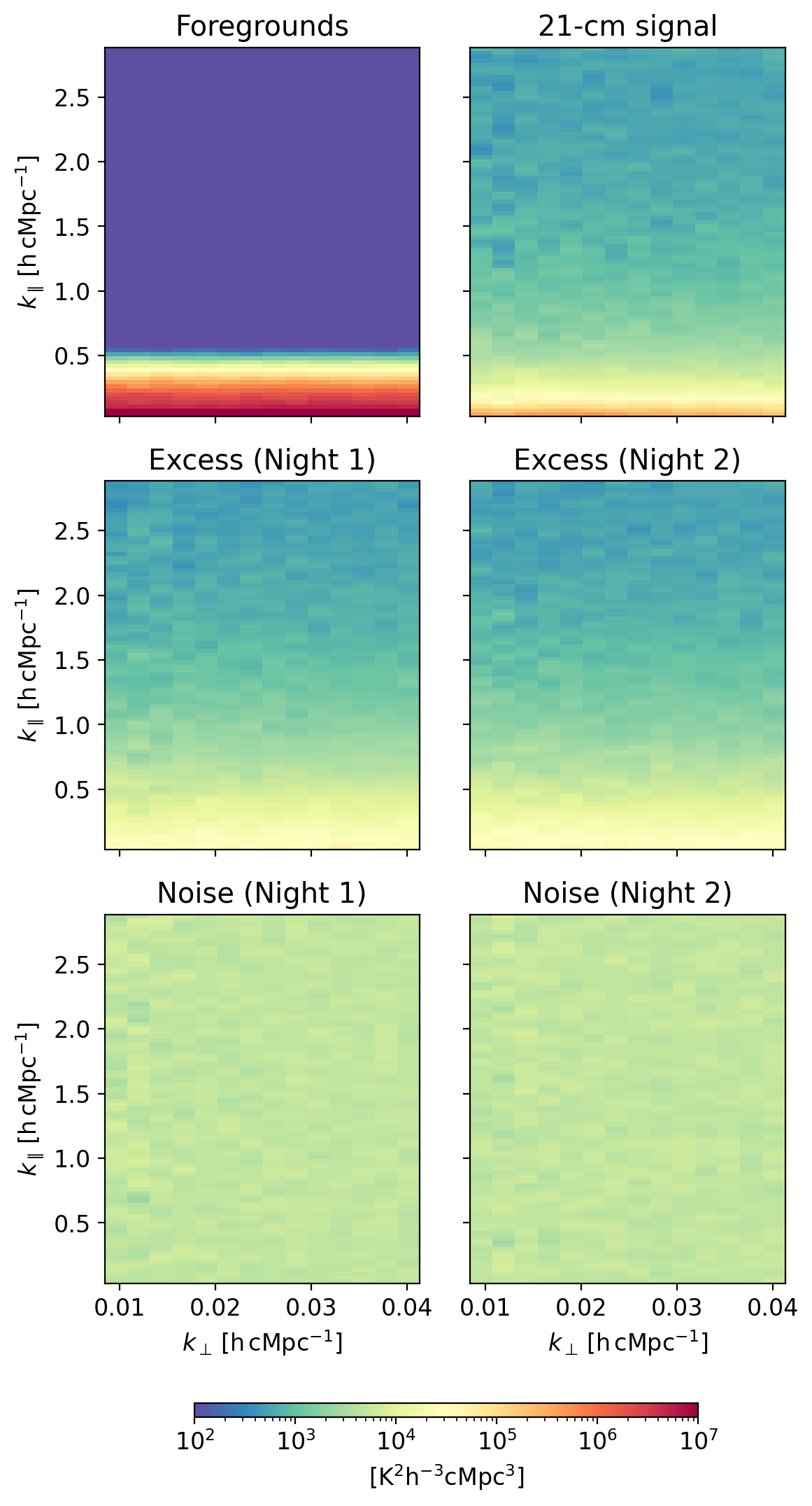}
    \caption{Cylindrical power spectra of the different input components constituting the simulated visibility cubes. The foreground and 21 cm signal components are coherent and have a common realisation for both nights. The excess and noise components are incoherent and have different realisations for the two nights.}
    \label{fig:inputs_sim}
\end{figure}

\begin{figure*}
\centering
    \includegraphics[width=1.8\columnwidth]{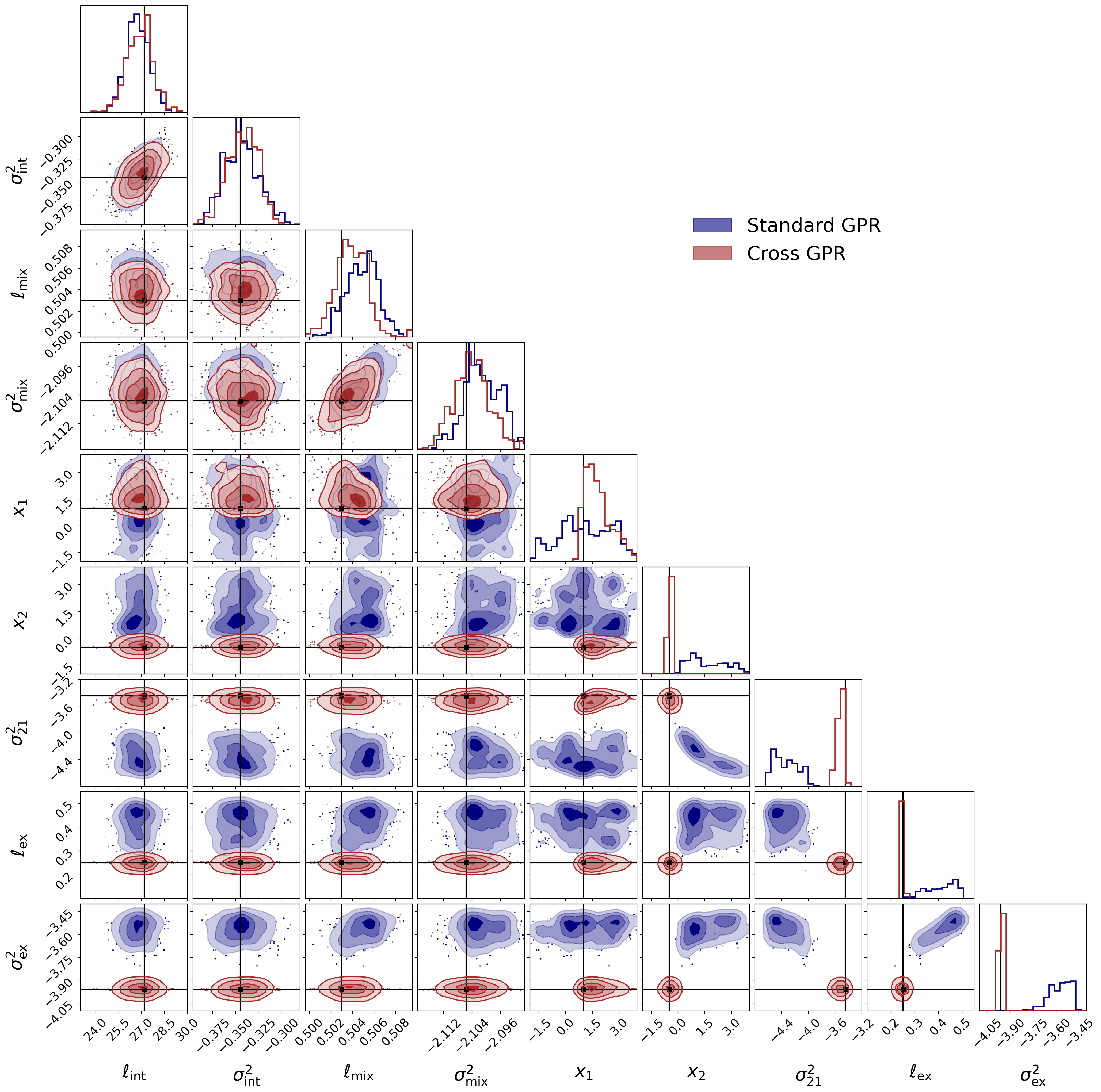}\vspace{0.5cm}
    \includegraphics[width=2\columnwidth]{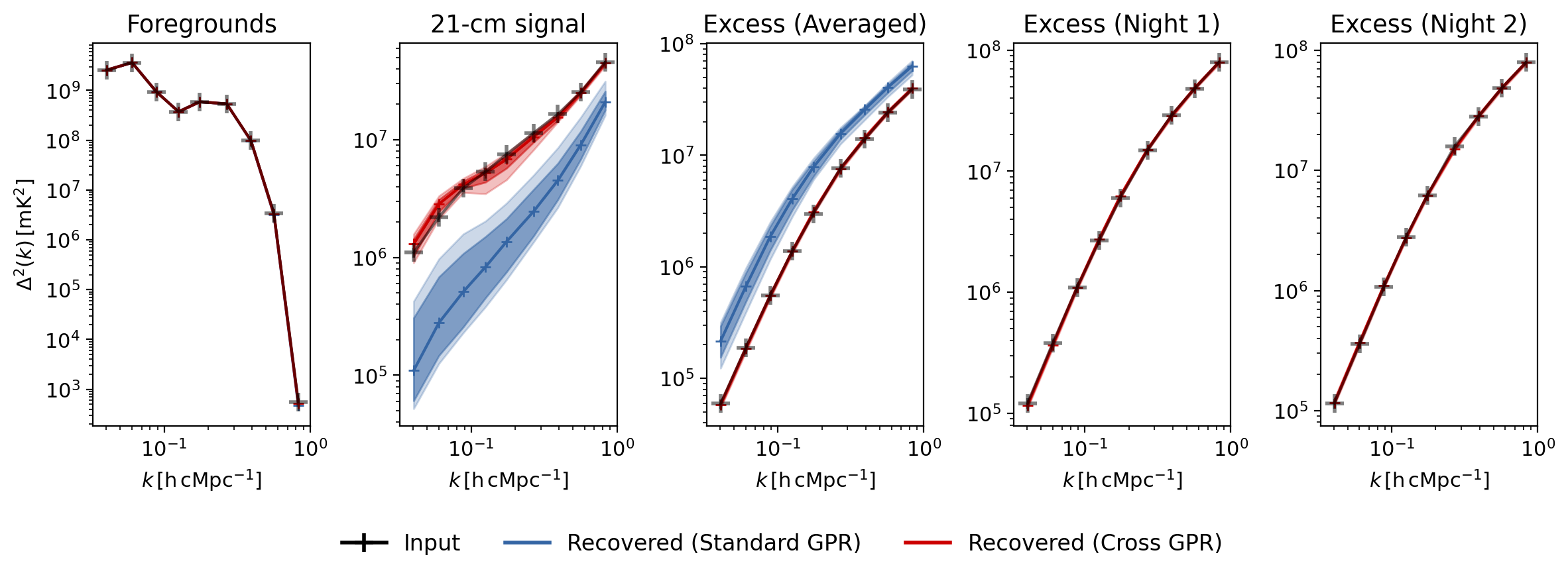}
    \caption{Comparison of the performance of standard and cross-GPR in recovering the input components. The standard and cross-GPR results are shown in blue and red colors, respectively. Top: Corner plot showing the posterior distribution of the GPR hyperparameters. The input hyperparameter values are indicated with black lines. Bottom: Input and recovered spherical power spectra of the different components.}
    \label{fig:corner_sim}
\end{figure*}

\section{Application to simulated data}\label{sec:simulations}
In this section, we demonstrate the cross-GPR approach on simulated visibility cubes. Such visibility cubes in $u\varv\nu$ space are obtained by gridding the visibility data obtained from radio interferometric observations \citep{offringa2019precision}. In the standard approach, gridded data cubes from multiple nights are combined by computing a weighted average (using the $u\varv\nu$ grid weights), and GPR is performed on this averaged data cube to subtract smooth spectrum foregrounds.

We note that performing two independent GPR runs on the two datasets, as described by Eq. (\ref{eq:joint_ind}), is not equivalent to performing a single GPR on the averaged data. Instead, throughout Sect. \ref{sec:joint_gpr_multiple}, we have considered the case where the GP model hyperparameters are linked between the two datasets, following Eq. (\ref{eq:joint_link}), and this is equivalent to a single frequency-only GPR performed on the averaged data. In Sect. \ref{sec:sim_21_excess}, we compare the cross-GPR approach against a frequency-only GPR performed on the averaged data. In Sect. \ref{sec:21cm_shape}, in addition to this, we also investigate the case of independent frequency-only GPR runs performed separately on the individual nights. We note that, unlike standard approaches used in LOFAR and NenuFAR analyses, where the excess component is retained, in our tests, the excess component was subtracted to assess its impact.

\subsection{Separating 21 cm signal and incoherent excess}\label{sec:sim_21_excess}
We consider a realistic scenario of an observation with the radio telescope NenuFAR. For this purpose, we simulated gridded visibility cubes of two nights of observations. The data cube contains coherent foreground power, which is made up of an intrinsic foreground component representing smooth spectrum foregrounds within the field of view and a mode-mixing component that represents off-axis source foreground power. We note that here the mode-mixing component is assumed to be night-to-night coherent if the two nights are observed in the same LST range. The excess variance component was assumed to be incoherent across the two nights. We included a 21 cm signal component with a variance equal to 0.1\% of the data variance, making sure it is above the thermal noise variance. This higher signal-to-noise ratio was used for illustration and comparison purposes only. The 21 cm visibility cube was generated from the VAE trained on 21 cm simulations at $z=20$ used by \cite{munshi2024first}. Finally, we simulated incoherent noise cubes for the two nights. For all components except the 21 cm signal component, the kernel function, and their associated lengthscale and variance values are inspired by those obtained by \cite{munshi2025improved} from actual data\footnote{\citet{munshi2025improved} used two Matern 3/2 kernels with different lengthscales to model the excess component. Here, for simplicity, we used a single Exponential kernel that exhibits a similar spectral structure to the combination of the two.}. The GP covariance model is summarised in Table \ref{tab:gpr}. We report all $\sigma^2$ parameters in $\log_{10}$ scale, and all $\ell$ parameters in linear scale. Power spectra were estimated from the visibility cubes using \texttt{ps\_eor}\footnote{\url{https://gitlab.com/flomertens/ps_eor}}. Figure~\ref{fig:inputs_sim} shows the cylindrical power spectra of the different components given as input to the simulation. The foreground component has high power at low $k_{\parallel}$ and has a sharp drop in power with increasing $k_{\parallel}$.\footnote{The baseline dependence of frequency-coherence scale introduced by \cite{mertens2024retrieving} has not been implemented in this paper, and is not relevant for the purpose of these simulations. This will be included in future simulations and applications.} The 21 cm and excess components are much weaker, and their power drop-off at high $k_{\parallel}$ is much slower. It is evident from the cylindrical power spectra that the 21 cm signal and excess component covariances are similar and might be difficult to disentangle based on spectral smoothness alone. The noise power for each night is approximately an order of magnitude below the peak excess power.

Given the simulated data, we next compare the performance of the cross-GPR approach developed in this paper to the more standard GPR approach, which only uses the frequency-frequency covariance within a single visibility cube. The averaged data and noise cubes obtained from these simulated visibility cubes for the two nights were given as input in the standard approach to sample the hyperparameter posterior distribution, and estimate the residuals and power spectra. We refer to this as the ``standard GPR'' case. Next, the data and noise cubes for both nights were given as input to the cross-GPR formalism developed in this paper to sample the hyperparameter posterior distribution, and estimate the residuals for each night, and corresponding power spectra for individual nights and averaged nights. We refer to this as the ``Cross-GPR'' case. In both cases, we used the same covariance functions, priors, and MCMC parameters. It was also ensured that the MCMC runs for both the standard and cross-GPR cases converged, in order to obtain an accurate representation of the posterior distribution in each case.

The top panel of Fig.~\ref{fig:corner_sim} shows the posterior distribution of the hyperparameters for standard GPR (in blue) and cross-GPR (in red). The recovery of the intrinsic ($\ell_\mathrm{int}$, $\sigma^2_\mathrm{int}$) and mode mixing ($\ell_\mathrm{mix}$, $\sigma^2_\mathrm{mix}$) foreground hyperparameters is comparable in both standard and cross-GPR. This is because they have distinct spectral behaviour, which is sufficient to disentangle them, as was seen in Sect.~\ref{sec:demo_synthetic}. We note, however, that the mode-mixing hyperparameter recovery, unlike in the cross-GPR case, is slightly biased in the standard GPR case. The remaining hyperparameters describe the 21 cm signal ($x_1$, $x_2$, $\sigma^2_{21}$) and the excess ($\ell_\mathrm{ex}$, $\sigma^2_\mathrm{ex}$). Here, cross-GPR performs dramatically better in recovering the input values compared to standard GPR. This could be expected since the excess and 21 cm signal components have similar spectral behavior (Fig.~\ref{fig:inputs_sim}) and are thus difficult to disentangle based on the spectral behavior alone, as seen in Sect.~\ref{sec:demo_synthetic} for similar kernels. Specifically, we find that in the standard GPR, $\sigma^2_{21}$ is underestimated while $\sigma^2_\mathrm{ex}$ is overestimated. This suggests that the excess component absorbs a portion of the 21 cm signal. The corner plot in the top panel of Fig.~\ref{fig:corner_sim} exhibits such a degeneracy between $\sigma^2_{21}$ and $\sigma^2_\mathrm{ex}$ and also between $\sigma^2_{21}$ and $x_2$. This is not the case in cross-GPR, where the recovery is unbiased. Thus, including information about the coherence or incoherence of components between nights through the cross-GPR approach can break the degeneracies between the hyperparameters, particularly for components with similar spectral behaviour, such as the 21 cm signal and excess components. We note that the cross-GPR method makes no additional assumptions beyond those of standard GPR, except for acknowledging that one component of the data is incoherent, which allows it to break degeneracies and more effectively separate the signals.

\begin{figure}
    \includegraphics[width=\columnwidth]{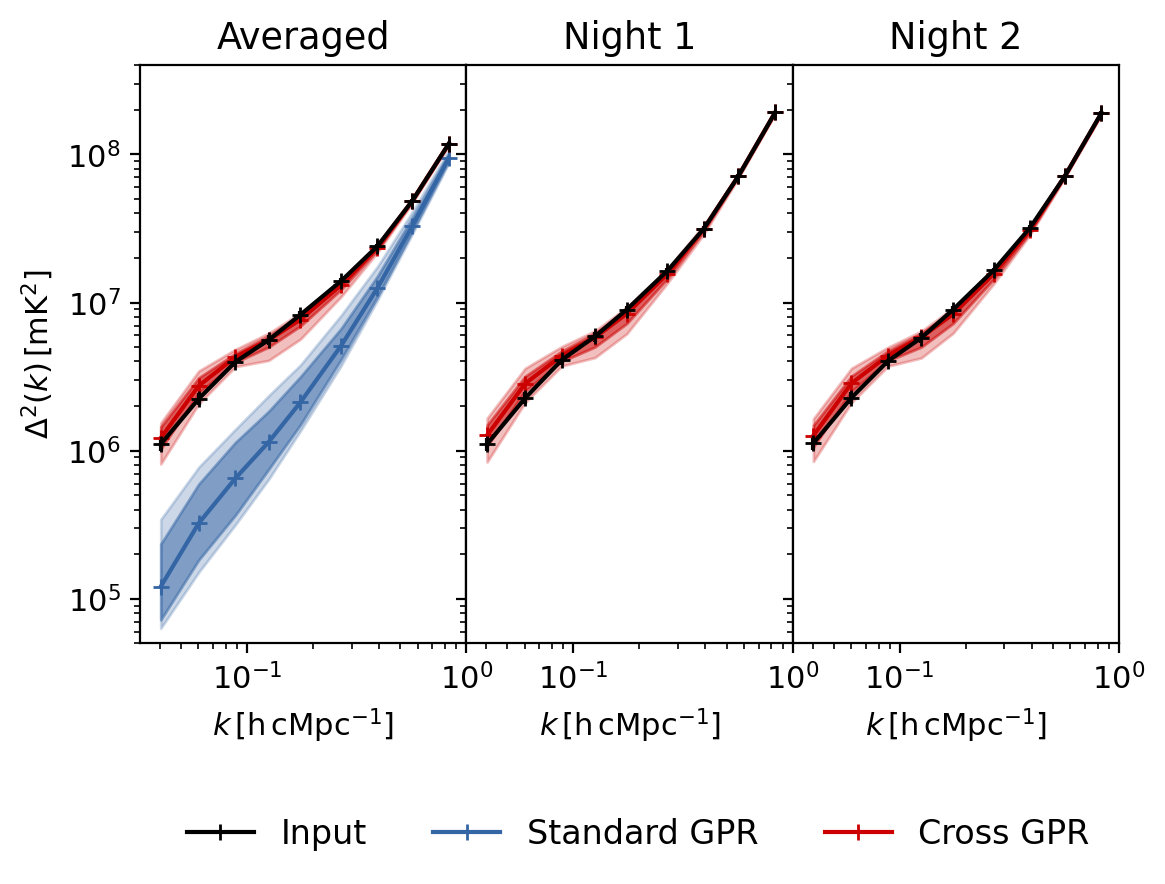}
    \caption{Power spectra of the residual data after the foregrounds and excess components have been subtracted. The standard and cross-GPR results are shown in blue and red colors, respectively.}
    \label{fig:subtract_sim}
\end{figure}
The bottom row of Fig.~\ref{fig:corner_sim} shows how well the different components are recovered by comparing their spherical power spectra. The shaded bands represent the $1\sigma$ and $2\sigma$ uncertainties that account for the spread of the hyperparameter posterior distribution. The foreground power spectrum is recovered by both the standard and cross-GPR approach relatively well (the blue and red lines are not clearly visible in the plot since they overlap). However, standard GPR underestimates the 21 cm signal component power and overestimates the excess power, because of the respective biases in their corresponding variances. Both the 21 cm signal and the excess component power spectra are recovered well in the cross-GPR approach. Cross-GPR allows us to estimate the recovered power spectra of incoherent components for each night, and this is shown in the two rightmost panels. The input power spectra are seen to be recovered well here as well.

Finally, we subtracted both the foreground and excess components from the data cubes that were used as the input to the GPR. We note that the upper limits analyses performed by \cite{mertens2020improved}, \cite{mertens2025deeper}, \cite{munshi2024first}, \cite{munshi2025improved}, and \cite{ceccotti2025first} using standard GPR do not subtract the excess component, because in the frequency-only approach, it cannot be safely disentangled from the 21 cm signal. For standard GPR, this was done on the averaged data. For cross-GPR, the components for each night obtained using Eq.~(\ref{eq:pred_fg_inc}) were subtracted from the respective nights and averaged to obtain the ensemble of averaged residual data cubes, from which the power spectra and uncertainties were estimated following Sect.~\ref{sec:joint_gpr_21cm}. Figure~\ref{fig:subtract_sim} shows the resulting spherical power spectra with uncertainties, compared with the respective input power spectra. A standard GPR oversubtracts the excess component and thus underestimates the residual power spectrum. The recovery in cross-GPR, on the other hand, is unbiased, in both the averaged data and in individual nights.

\subsection{Dependence on 21 cm signal power spectrum shapes}\label{sec:21cm_shape}
In this section, we assess how the recovery of the hyperparameter distributions and the power spectra depends on the degree of similarity between the shape of the 21 cm signal and the excess component power spectra. We repeated the simulations described in the previous section for a series of 21 cm signal shapes. We sampled 25 different shapes of the signal from the trained VAE, corresponding to $5\times 5$ grid of uniformly distributed points in the two-dimensional latent space ($x_1,x_2\in [-2,2]$). For each such signal shape, a 21 cm signal visibility cube was generated and added to the coherent foregrounds visibility cube and incoherent excess and noise cubes for both nights. The standard and cross-GPR were then run in the same manner as described in Sect.~\ref{sec:sim_21_excess}. In addition to running standard GPR on the averaged data and cross-GPR on the set of two nights, here we also investigated the case where standard GPR is run on each of the nights separately.

\begin{figure}
    \includegraphics[width=\columnwidth]{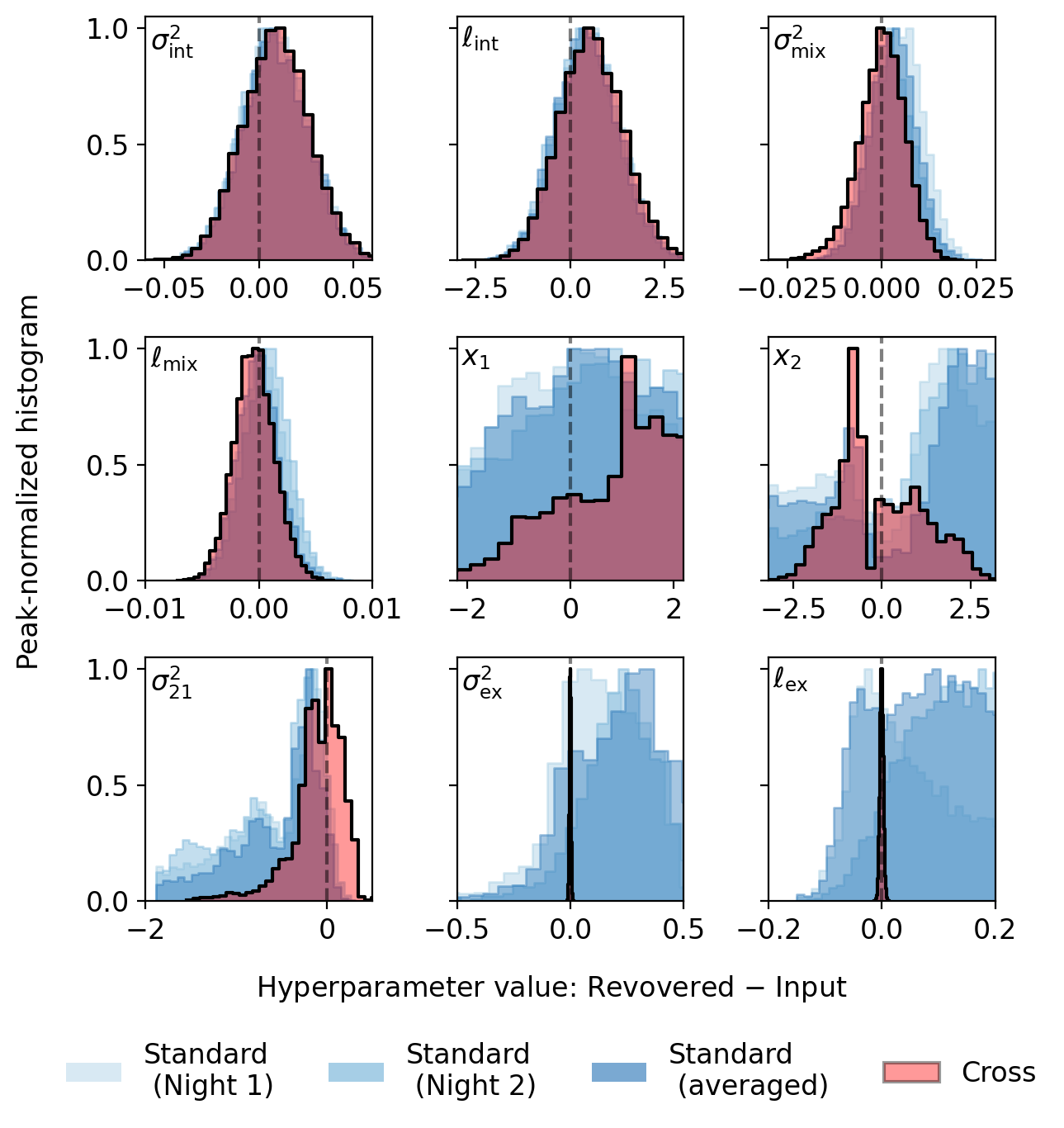}
    \caption{Recovery of the input hyperparameters for simulations performed for a variety of 21 cm signal shapes. The peak-normalised histograms marginalised over all signal shapes, computed after subtracting the input values, are shown. The different panels correspond to the different hyperparameters. The results from the three runs of standard GPR are shown in different shades of blue, while the cross-GPR results are shown in red. A vertical line is plotted at Recovered = Input to indicate perfect recovery.}
    \label{fig:rec_params}
\end{figure}

In Fig.~\ref{fig:rec_params}, we show the hyperparameter recovery results, in the form of peak normalised histograms for each hyperparameter marginalised over all 25 simulations. The input true value of the hyperparameter was subtracted before computing the histogram for each signal shape so that it is centered at zero for unbiased recovery. The histograms for the three sets of standard GPR runs are shown in different shades of blue, while the histogram for cross-GPR hyperparameter recovery is shown in red. We find that the foreground hyperparameters are relatively well recovered by both standard and cross-GPR, agreeing with the results of Sect.~\ref{sec:sim_21_excess}. Also, similar to what is seen in Fig.~\ref{fig:corner_sim}, the mode-mixing foreground variance is slightly less biased in the cross-GPR approach. However, we find that the $x_1$ parameter is poorly recovered in both the standard and cross-GPR approaches, likely due to the weak dependence of the trained VAE on $x_1$. This was also observed in Fig. \ref{fig:corner_sim}, where the range of $x_1$ is much larger than that of $x_2$. The recovery of $x_2$ is also suboptimal in both cases, although the cross-GPR results appear slightly better constrained around zero. The 21 cm signal variance recovery is, however, significantly improved in the cross-GPR approach, where the standard GPR underestimates the 21 cm signal variance, indicating that a portion of the 21 cm signal is absorbed by the excess component. The most significant improvement is seen in the recovery of both the lengthscale and variance of the incoherent excess component. The cross-GPR approach accurately recovers the input values, while the standard GPR results remain largely unconstraining. This is expected, as the explicit information on the nature of the variance provided to the model is precisely the incoherence of the excess component.

\begin{figure*}
    \includegraphics[width=2\columnwidth]{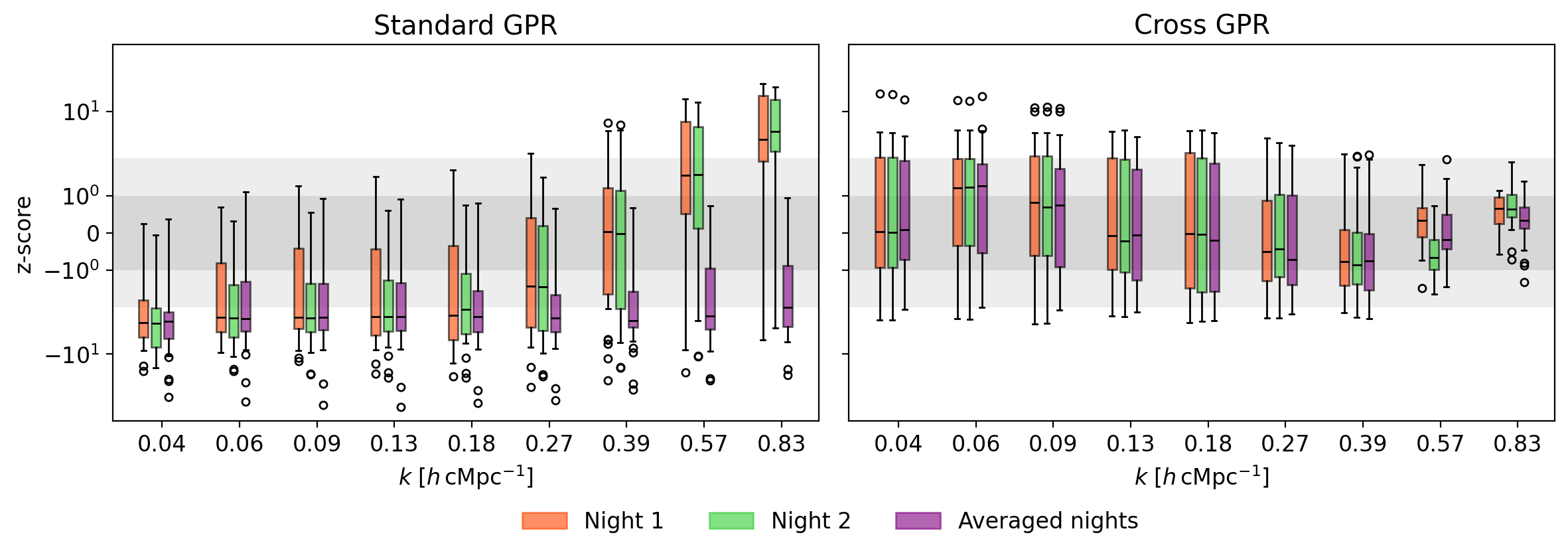}\vspace{0.2cm}
    \includegraphics[width=2\columnwidth]{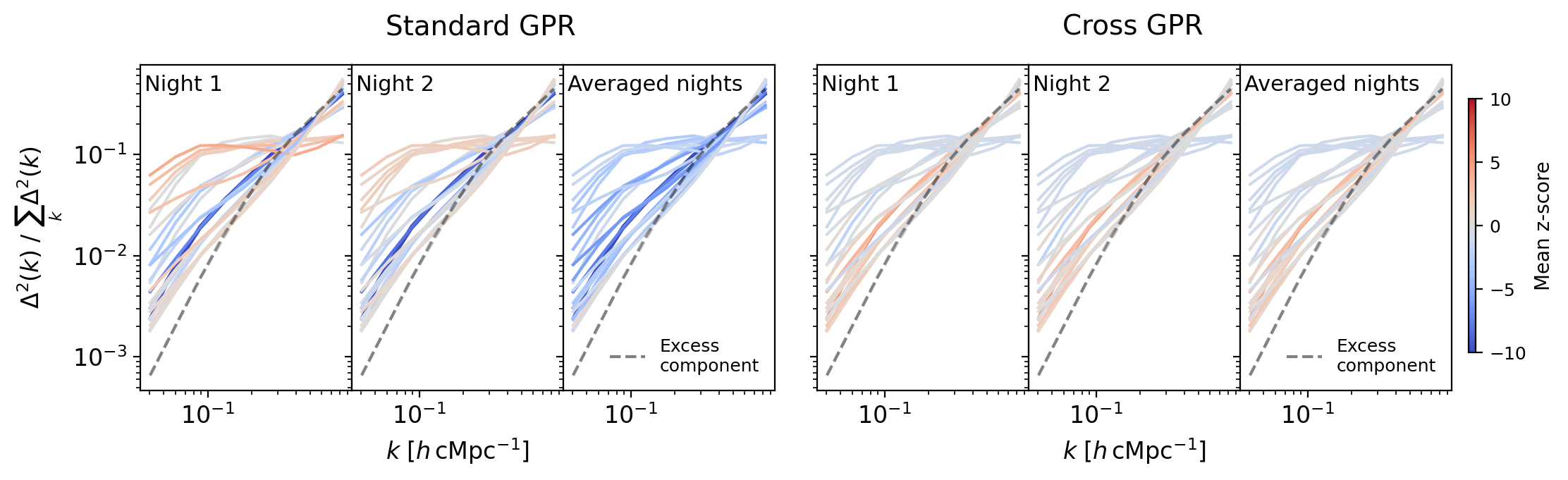}
    \caption{Distribution of z-scores for simulations performed for a variety of 21 cm signal shapes. The two columns show the results for standard and cross-GPR. The top panel shows the dependence of the z-score on $k$, for the different nights. The gray bands indicate the $1\sigma$ and $2\sigma$ levels. The bottom panel shows the dependence of the mean z-score on the 21 cm signal shape. The normalised power spectra for the different signal shapes are shown with colors indicating the mean z-score. The shape of the excess component power spectrum is indicated in each panel with  dashed black lines.}
    \label{fig:z_scores}
\end{figure*}
Once the hyperparameter posterior distribution was obtained, the foregrounds and excess were subtracted from the input data, and the residual power spectrum was estimated in the manner described in Sect.~\ref{sec:sim_21_excess}, for each signal shape. Finally, a bias correction of the thermal noise was made by subtracting the thermal noise power spectrum. The resulting power spectrum uncertainties should bracket the input 21 cm power spectrum for unbiased recovery. To estimate this, we computed the z-score value as a function of $k$ for each simulation, defined as
\begin{equation}
    \text{z-score}(k) = \dfrac{\Delta^2_\mathrm{rec}(k)-\Delta^2_\mathrm{inp}(k)}{\sigma_{\Delta^2_\mathrm{rec}}(k)},
\end{equation}
where $\Delta^2_\mathrm{rec}(k)$ and $\sigma_{\Delta^2_\mathrm{rec}}(k)$ are the residual noise bias corrected power spectrum and 1$\sigma$ uncertainty, while $\Delta^2_\mathrm{inp}(k)$ is the input 21 cm signal power spectrum. A negative value of the z-score indicates 21 cm signal loss. The z-scores computed in this manner are shown in Fig.~\ref{fig:z_scores}. The top row shows the dependence of the z-score on $k$. Each box describes the spread of z-scores for the 25 selected input signal shapes. The gray bands indicate the $1\sigma$ and $2\sigma$ ranges. We find that the standard GPR residuals systematically underestimate the 21 cm signal at most $k$ modes, with a large number of z-score values below $-2$, indicating suppression of the signal beyond the $2\sigma$ level. At some large $k$ modes, for GPR performed on individual nights, a significant overestimation is also seen with z-score values well above 2, hence the bias is scale-dependent. The cross-GPR does not show such a signal loss, with the recovery at most $k$ modes being well constrained in the $2\sigma$ bands independent of the $k$ mode. A few simulations show a positive bias at small $k$, but it is still lower than the standard GPR results, and such outliers can be expected given the number of simulations. We note that such an overestimation is not as concerning as an underestimation, particularly in the context of setting upper limits on the power spectrum.

To more precisely investigate the dependence of the signal recovery on the input 21 cm signal shape, the bottom panel of Fig.~\ref{fig:z_scores} illustrates the dependence of the z-score on the signal shape. For each signal shape, we computed the mean z-score over $k$. The normalised power spectrum representing each signal shape is shown with its color indicating the corresponding value of the mean z-score. As a reference, we also show a normalised power spectrum shape of the excess component as  dashed black lines. In standard GPR, a large number of signal shapes are significantly suppressed. We find that the signals that are the most suppressed have a power spectrum shape similar to the reference excess component. This supports the findings of Sects. \ref{sec:demo_synthetic} and \ref{sec:sim_21_excess}, which show that signals similar to the excess are more difficult to recover in the standard approach. In contrast, the cross-GPR approach more effectively separates such signals. In fact, the trend appears reversed: signals resembling the excess component show a slight positive bias, suggesting that the 21 cm component absorbs part of the excess. However, a slight positive bias is less concerning as long as we are in the regime of setting upper limits on the 21 cm power spectrum, since the signal is not absorbed. None of the signals show a significant negative bias, with all z-scores closer to zero and lying within the $2\sigma$ bounds. The absolute values of the z-score, averaged over both $k$ bins and shapes of the signal are 4.6, 4.8, and 5.7 for night 1, night 2, and the averaged night simulations, respectively, with a standard GPR. The corresponding values obtained with cross-GPR are 1.6, 1.6, and 1.5, indicating a significant decrease in the absolute bias.

\section{Discussion}\label{sec:discussion}
In this section, we outline the key features of the novel cross-GPR method developed in this paper and assess its current limitations. We also discuss its implications for future analyses in 21 cm cosmology.

\subsection{Main features}
We developed a generalised framework that explicitly incorporates (in)coherence between specific signal components across multiple datasets to disentangle them using GPR. Our method extends the single-dataset GPR formulation by introducing cross-covariances between multiple datasets to encode whether a given component is coherent or incoherent. Our simulations demonstrate that this additional dimension significantly improves our ability to separate signals with similar spectral behavior.

This is particularly relevant in 21 cm cosmology, where signal recovery is complicated due to the presence of foregrounds and various systematics, some of which might not exhibit the spectral smoothness that is required by standard frequency-only GPR approaches to separate them from the 21 cm signal. In the context of 21 cm cosmology, the multiple datasets in the cross-GPR approach correspond to different nights of observations, where the foregrounds and 21 cm signal are coherent across nights, but a portion of the contributions to the excess variance are not. The night-to-night (in)coherence used in this method effectively provides an additional signal subspace, orthogonal to spectral smoothness. Our simulations show that this cross-GPR approach can break degeneracies between the coherent 21 cm signal and the incoherent excess variance. This enables us to more confidently subtract incoherent contributions to the excess variance, without suppressing the cosmological 21 cm signal. The framework retains the flexibility of standard GPR, including the use of physically informed kernels and machine-learned 21 cm signal covariances, while adding another discriminator, the night-to-night (in)coherence.

\subsection{Implications for 21 cm cosmology analyses}
Multiple interferometers attempting a statistical detection of the 21 cm signal suffer from excess variance above the thermal noise, which typically has a small frequency coherence scale that cannot be reliably separated from the 21 cm signal. Even though the incoherent portion of this excess variance integrates down as multiple nights of observation are averaged, it prevents us from exploiting the full sensitivity of the instrument if left in the data. This framework provides a robust algorithm for subtracting such incoherent contributions to the excess variance. Such a subtraction removes modeled realisations of the incoherent component (and foregrounds) from the data itself, thus reducing both its bias and sample variance. If the excess variance after foreground subtraction is primarily incoherent, this approach can significantly improve the sensitivity to the 21 cm signal by bringing the residuals closer to the thermal noise sensitivity. Additionally, the method offers a new diagnostic that enables us to analyze the coherence properties of residuals in visibility cubes across nights. This can provide additional insights into the nature of the dominant systematics in a given instrument. 

Building on the frequency-frequency GPR framework already in use for instruments such as LOFAR and NenuFAR, this method operates on gridded visibility cubes, making it directly applicable to any phase tracking interferometer such as LOFAR, NenuFAR, MWA, and the upcoming SKA\footnote{Square Kilometre Array} \citep{dewdney2009square}, which employ the reconstructed power spectrum approach. Additionally, this method is general and can also be extended to HI intensity mapping experiments, or drift-scan 21 cm cosmology instruments such as HERA, which employ the delay spectrum approach to estimate the 21 cm power spectrum \citep{parsons2012per}. For delay spectrum approaches, cross-GPR could extend the GPR applications on calibrated visibilities from a given LST bin, as performed by \cite{ghosh2020foreground}, to data from multiple nights at the same LST or across sets of redundant baselines. A key next step is thus to apply this to real observational data, which is particularly straightforward for instruments such as LOFAR and NenuFAR, which already use standard GPR in their data processing pipelines. This could lead to deeper and robust upper limits on the 21 cm signal power spectrum with these instruments. 

\subsection{Limitations and future improvement}
The foremost limitation of the method is the computational scalability to a large number of nights. As discussed in Sect.~\ref{sec:generalization}, a direct generalisation from two to $N$ nights is computationally expensive, since the complexity of the inversion of the joint covariance matrix involved in the likelihood evaluation grows as $\mathcal{O}(N^3)$. One possibility to avoid this, also described in Sect.~\ref{sec:generalization}, is to pair each night with one other night, resulting in a block diagonal covariance matrix consisting of $N/2$ independent blocks. This allows the use of sparse matrix algorithms for inversion, but captures (in)coherence only between each pair. Intermediate approaches are also possible, for instance, using blocks of three nights, or constructing a banded covariance matrix (for example, tridiagonal), which would incorporate limited coherence information across neighboring nights while still remaining computationally feasible. Another possibility is to divide the data into two sets of nights and average each set separately. The averaged visibility cube in each set could then be used by the algorithm to identify and subtract the incoherent component. In such cases, randomising the set of nights that go into each averaged dataset could lead to similar covariance matrices, because of the central limit theorem.

In our simulations, we assumed that the foregrounds were completely coherent across nights. This assumption holds exactly if the $u\varv$ coverage for the different nights is the same; however, this might not be the case, for example, if the data flagging were to be significantly different across nights. Differences in $u\varv$ coverage can induce artificial decoherence, but this is expected to be minor for directions near the phase center, such as the 21 cm signal and foregrounds within the primary beam main lobe. A small portion of the mode-mixing foregrounds, however, can become incoherent, and the foreground model would then need to account for smooth-frequency incoherent contributions to the data power. Such a split of the foreground kernel into coherent and incoherent components increases the number of hyperparameters being optimised, thereby raising the computational cost. We note that increasing the number of components can increase the propagated error bars on the 21 cm power spectrum; however,  this might be counteracted by the fact that an improved GP model reduces the degeneracies and thus decreases the spread of the hyperparameter distribution, in addition to reducing biases. So the uncertainties can decrease even if the number of components increases. Additionally, any effect that causes decoherence of the 21 cm signal itself across multiple nights needs to be carefully investigated, and its impact on the method needs to be assessed. Cross-coherence analyses by \cite{mertens2020improved,mertens2025deeper,munshi2025improved} show that for LOFAR and NenuFAR, the signal at low $k_{\parallel}$, dominated by foreground emission in the main lobe of the primary beam, exhibits a coherence that is very close to unity across multiple nights at similar LST. This suggests that decoherence due to non-overlapping $u\varv$ sampling is at most second order, but the level of decoherence could still exceed the level of the incoherent excess.

Finally, we assumed that all nights had the same covariance structure with shared hyperparameters. Both these assumptions can be lifted in the future. Allowing for different hyperparameters for the incoherent component and, ultimately, allowing for different covariance kernels for the incoherent component for different nights could be necessary to model visibility data across a large number of nights, which could suffer from the effects of very different sources for the excess variance.

\section{Conclusions}\label{sec:conclusion}
In this paper, we present a novel framework, based on GPR, which uses information about the coherence of specific components across multiple datasets to disentangle them. The method is a good fit to signal separation problems in 21 cm cosmology, allowing the coherent 21 cm signal to be separated from time-incoherent systematic contributions to the excess variance above the thermal noise. The main conclusions from this analysis are summarised below.

Simulations using general synthetic data as well as radio interferometric visibility cubes show that this framework can break the degeneracy between signal components in situations where standard GPR approaches fail to do so. The improvement is particularly pronounced when signal components have similar spectral behaviors, but different coherence behaviors across datasets. Simulations for a wide variety of 21 cm signals show that the incoherent contributions to the excess variance modeled using this approach could be subtracted from the data without suppressing the underlying 21 cm signal. Such a reliable subtraction of a portion of the excess variance could improve the achieved sensitivity of 21 cm experiments. Our approach builds on the standard frequency-frequency GPR framework, already in use for LOFAR and NenuFAR and operating on gridded visibility cubes, and extends it by incorporating cross-dataset covariance terms. Applications of the method to real data from interferometers such as LOFAR, NenuFAR, MWA, and the upcoming SKA are, thus, straightforward and could enable more stringent upper limits on the 21 cm signal power spectrum using current observations. The implementation is publicly available as part of the Python library \texttt{crossgp}\footnote{\url{https://github.com/satyapan/crossgp}} and will soon be integrated into the LOFAR and NenuFAR foreground removal and power spectrum estimation framework \texttt{ps\_eor}.

\begin{acknowledgements}
SM and LVEK acknowledge the financial support from the European Research Council (ERC) under the European Union’s Horizon 2020 research and innovation programme (Grant agreement No. 884760, "CoDEX”). FGM acknowledges support from the I-DAWN project, funded by the DIM-ORIGINS programme. EC would like to acknowledge support from the Centre for Data Science and Systems Complexity (DSSC), Faculty of Science and Engineering at the University of Groningen, and from the Ministry of Universities and Research (MUR) through the PRIN project `Optimal inference from radio images of the epoch of reionisation'.
\end{acknowledgements}

\bibliographystyle{aa}
\bibliography{aa}

@article{adams2023improved,
  title={Improved constraints on the 21 cm EoR power spectrum and the X-ray heating of the IGM with HERA phase I observations},
  author={Adams, Tyrone and Aguirre, James E and Alexander, Paul and Ali, Zaki S and Baartman, Rushelle and Balfour, Yanga and Barkana, Rennan and Beardsley, Adam P and Bernardi, Gianni and Billings, Tashalee S and others},
  journal={ApJ},
  volume={945},
  number={2},
  pages={124},
  year={2023},
  publisher={IOP Publishing}
}

@article{mertens2025deeper,
  title={Deeper multi-redshift upper limits on the epoch of reionisation 21 cm signal power spectrum from LOFAR between z= 8.3 and z= 10.1},
  author={Mertens, FG and Mevius, M and Koopmans, LVE and Offringa, AR and Zaroubi, S and Acharya, A and Brackenhoff, SA and Ceccotti, E and Chapman, E and Chege, K and others},
  journal={A\&A},
  volume={698},
  pages={A186},
  year={2025},
  publisher={EDP Sciences}
}

@article{paciga2013simulation,
  title={A simulation-calibrated limit on the H i power spectrum from the GMRT Epoch of Reionization experiment},
  author={Paciga, Gregory and Albert, Joshua G and Bandura, Kevin and Chang, Tzu-Ching and Gupta, Yashwant and Hirata, Christopher and Odegova, Julia and Pen, Ue-Li and Peterson, Jeffrey B and Roy, Jayanta and others},
  journal={MNRAS},
  volume={433},
  number={1},
  pages={639--647},
  year={2013},
  publisher={The Royal Astronomical Society}
}

@article{barry2019improving,
  title={Improving the epoch of reionization power spectrum results from Murchison Widefield Array season 1 observations},
  author={Barry, N and Wilensky, M and Trott, CM and Pindor, B and Beardsley, AP and Hazelton, BJ and Sullivan, IS and Morales, MF and Pober, JC and Line, J and others},
  journal={ApJ},
  volume={884},
  number={1},
  pages={1},
  year={2019},
  publisher={IOP Publishing}
}

@article{li2019first,
  title={First season MWA Phase II epoch of reionization power spectrum results at redshift 7},
  author={Li, W and Pober, JC and Barry, N and Hazelton, BJ and Morales, MF and Trott, CM and Lanman, A and Wilensky, M and Sullivan, I and Beardsley, AP and others},
  journal={ApJ},
  volume={887},
  number={2},
  pages={141},
  year={2019},
  publisher={IOP Publishing}
}

@article{trott2020deep,
  title={Deep multiredshift limits on Epoch of Reionization 21 cm power spectra from four seasons of Murchison Widefield Array observations},
  author={Trott, Cathryn M and Jordan, CH and Midgley, S and Barry, N and Greig, B and Pindor, B and Cook, JH and Sleap, G and Tingay, SJ and Ung, D and others},
  journal={MNRAS},
  volume={493},
  number={4},
  pages={4711--4727},
  year={2020},
  publisher={Oxford University Press}
}

@article{kolopanis2019simplified,
  title={A simplified, lossless reanalysis of PAPER-64},
  author={Kolopanis, Matthew and Jacobs, Daniel C and Cheng, Carina and Parsons, Aaron R and Kohn, Saul A and Pober, Jonathan C and Aguirre, James E and Ali, Zaki S and Bernardi, Gianni and Bradley, Richard F and others},
  journal={ApJ},
  volume={883},
  number={2},
  pages={133},
  year={2019},
  publisher={IOP Publishing}
}

@article{patil2017upper,
  title={Upper limits on the 21 cm epoch of reionization power spectrum from one night with LOFAR},
  author={Patil, AH and Yatawatta, S and Koopmans, LVE and De Bruyn, AG and Brentjens, MA and Zaroubi, S and Asad, KMB and Hatef, M and Jeli{\'c}, Vibor and Mevius, M and others},
  journal={ApJ},
  volume={838},
  number={1},
  pages={65},
  year={2017},
  publisher={IOP Publishing}
}

@article{mertens2020improved,
  title={Improved upper limits on the 21 cm signal power spectrum of neutral hydrogen at z≈ 9.1 from LOFAR},
  author={Mertens, Florent G and Mevius, M and Koopmans, Leon VE and Offringa, AR and Mellema, Garrelt and Zaroubi, Saleem and Brentjens, MA and Gan, H and Gehlot, Bharat Kumar and Pandey, VN and others},
  journal={MNRAS},
  volume={493},
  number={2},
  pages={1662--1685},
  year={2020},
  publisher={Oxford University Press}
}

@article{abdurashidova2022first,
  title={First results from hera phase i: Upper limits on the epoch of reionization 21 cm power spectrum},
  author={Abdurashidova, Zara and Aguirre, James E and Alexander, Paul and Ali, Zaki S and Balfour, Yanga and Beardsley, Adam P and Bernardi, Gianni and Billings, Tashalee S and Bowman, Judd D and Bradley, Richard F and others},
  journal={ApJ},
  volume={925},
  number={2},
  pages={221},
  year={2022},
  publisher={IOP Publishing}
}

@article{eastwood201921,
  title={The 21 cm power spectrum from the cosmic dawn: first results from the OVRO-LWA},
  author={Eastwood, Michael W and Anderson, Marin M and Monroe, Ryan M and Hallinan, Gregg and Catha, Morgan and Dowell, Jayce and Garsden, Hugh and Greenhill, Lincoln J and Hicks, Brian C and Kocz, Jonathon and others},
  journal={AJ},
  volume={158},
  number={2},
  pages={84},
  year={2019},
  publisher={IOP Publishing}
}

@article{gehlot2019first,
  title={The first power spectrum limit on the 21-cm signal of neutral hydrogen during the Cosmic Dawn at z= 20--25 from LOFAR},
  author={Gehlot, BK and Mertens, FG and Koopmans, LVE and Brentjens, MA and Zaroubi, S and Ciardi, B and Ghosh, A and Hatef, M and Iliev, Ilian T and Jeli{\'c}, V and others},
  journal={MNRAS},
  volume={488},
  number={3},
  pages={4271--4287},
  year={2019},
  publisher={Oxford University Press}
}

@article{ewall2016first,
  title={First limits on the 21 cm power spectrum during the Epoch of X-ray heating},
  author={Ewall-Wice, Aaron and Dillon, Joshua S and Hewitt, Jacqueline N and Loeb, Abraham and Mesinger, A and Neben, AR and Offringa, AR and Tegmark, Max and Barry, N and Beardsley, AP and others},
  journal={MNRAS},
  volume={460},
  number={4},
  pages={4320--4347},
  year={2016},
  publisher={Oxford University Press}
}

@article{garsden202121,
  title={A 21-cm power spectrum at 48 MHz, using the Owens Valley Long Wavelength Array},
  author={Garsden, Hugh and Greenhill, Lincoln and Bernardi, Gianni and Fialkov, Anastasia and Price, Daniel C and Mitchell, Daniel and Dowell, Jayce and Spinelli, Marta and Schinzel, Frank K},
  journal={MNRAS},
  volume={506},
  number={4},
  pages={5802--5817},
  year={2021},
  publisher={Oxford University Press}
}

@article{gehlot2020aartfaac,
  title={The AARTFAAC Cosmic Explorer: observations of the 21-cm power spectrum in the EDGES absorption trough},
  author={Gehlot, BK and Mertens, FG and Koopmans, LVE and Offringa, AR and Shulevski, A and Mevius, M and Brentjens, MA and Kuiack, M and Pandey, VN and Rowlinson, A and others},
  journal={MNRAS},
  volume={499},
  number={3},
  pages={4158--4173},
  year={2020},
  publisher={Oxford University Press}
}

@article{yoshiura2021new,
  title={A new MWA limit on the 21 cm power spectrum at redshifts~ 13--17},
  author={Yoshiura, S and Pindor, B and Line, JLB and Barry, Nichole and Trott, CM and Beardsley, A and Bowman, J and Byrne, R and Chokshi, A and Hazelton, BJ and others},
  journal={MNRAS},
  volume={505},
  number={4},
  pages={4775--4790},
  year={2021},
  publisher={Oxford University Press}
}

@article{munshi2024first,
  title={First upper limits on the 21 cm signal power spectrum from cosmic dawn from one night of observations with NenuFAR},
  author={Munshi, S and Mertens, FG and Koopmans, LVE and Offringa, AR and Semelin, B and Aubert, D and Barkana, R and Bracco, A and Brackenhoff, SA and Cecconi, B and others},
  journal={A\&A},
  volume={681},
  pages={A62},
  year={2024},
  publisher={EDP Sciences}
}

@ARTICLE{nunhokee2025limits,
       author = {{Nunhokee}, C.~D. and {Null}, D. and {Trott}, C.~M. and {Barry}, N. and {Qin}, Y. and {Wayth}, R.~B. and {Line}, J.~L.~B. and {Jordan}, C.~H. and {Pindor}, B. and {Cook}, J.~H. and {Bowman}, J. and {Chokshi}, A. and {Ducharme}, J. and {Elder}, K. and {Guo}, Q. and {Hazelton}, B. and {Hidayat}, W. and {Ito}, T. and {Jacobs}, D. and {Jong}, E. and {Kolopanis}, M. and {Kunicki}, T. and {Lilleskov}, E. and {Morales}, M.~F. and {Pober}, J.~C. and {Selvaraj}, A. and {Shi}, R. and {Takahashi}, K. and {Tingay}, S.~J. and {Webster}, R.~L. and {Yoshiura}, S. and {Zheng}, Q.},
        title = "{Limits on the 21 cm Power Spectrum at z = 6.5{\textendash}7.0 from Murchison Widefield Array Observations}",
      journal = {\apj},
         year = 2025,
        month = aug,
       volume = {989},
       number = {1},
          eid = {57},
        pages = {57},
          doi = {10.3847/1538-4357/adda45},
archivePrefix = {arXiv},
       eprint = {2505.09097},
 primaryClass = {astro-ph.CO},
       adsurl = {https://ui.adsabs.harvard.edu/abs/2025ApJ...989...57N}
}

@article{brackenhoff2025robust,
  title={Robust direction-dependent gain-calibration of beam-modelling errors far from the target field},
  author={Brackenhoff, SA and Offringa, AR and Mevius, M and Koopmans, LVE and Chege, JK and Ceccotti, E and H{\"o}fer, C and Gao, L and Ghosh, S and Mertens, FG and others},
  journal={MNRAS},
  pages={staf1206},
  year={2025},
  publisher={Oxford University Press}
}

@article{chokshi2024necessity,
  title={The necessity of individually validated beam models for an interferometric epoch of reionization detection},
  author={Chokshi, A and Barry, N and Line, JLB and Jordan, CH and Pindor, B and Webster, RL},
  journal={MNRAS},
  volume={534},
  number={3},
  pages={2475--2484},
  year={2024},
  publisher={Oxford University Press}
}

@article{kern2019mitigating,
  title={Mitigating internal instrument coupling for 21 cm cosmology. I. Temporal and spectral modeling in simulations},
  author={Kern, Nicholas S and Parsons, Aaron R and Dillon, Joshua S and Lanman, Adam E and Fagnoni, Nicolas and de Lera Acedo, Eloy},
  journal={ApJ},
  volume={884},
  number={2},
  pages={105},
  year={2019},
  publisher={IOP Publishing}
}

@article{kern2020mitigating,
  title={Mitigating internal instrument coupling for 21 cm cosmology. II. A method demonstration with the Hydrogen Epoch of Reionization Array},
  author={Kern, Nicholas S and Parsons, Aaron R and Dillon, Joshua S and Lanman, Adam E and Liu, Adrian and Bull, Philip and Ewall-Wice, Aaron and Abdurashidova, Zara and Aguirre, James E and Alexander, Paul and others},
  journal={ApJ},
  volume={888},
  number={2},
  pages={70},
  year={2020},
  publisher={IOP Publishing}
}

@article{mertens2018statistical,
  title={Statistical 21-cm signal separation via Gaussian Process Regression analysis},
  author={Mertens, FG and Ghosh, A and Koopmans, LVE},
  journal={MNRAS},
  volume={478},
  number={3},
  pages={3640--3652},
  year={2018},
  publisher={Oxford University Press}
}

@book{Rasmussen2006Gaussian,
  author = {Rasmussen, Carl Edward and Williams, Christopher K. I.},
  publisher = {The MIT Press},
  title = {Gaussian Processes for Machine Learning},
  year = 2006
}

@article{alvarez2012kernels,
  title={Kernels for vector-valued functions: A review},
  author={Alvarez, Mauricio A and Rosasco, Lorenzo and Lawrence, Neil D and others},
  journal={Foundations and Trends{\textregistered} in Machine Learning},
  volume={4},
  number={3},
  pages={195--266},
  year={2012},
  publisher={Now Publishers, Inc.}
}

@book{goovaerts1997geostatistics,
  title={Geostatistics for natural resources evaluation},
  author={Goovaerts, Pierre},
  year={1997},
  publisher={Oxford university press}
}

@inproceedings{teh2005semiparametric,
  title={Semiparametric latent factor models},
  author={Teh, Yee Whye and Seeger, Matthias and Jordan, Michael I},
  booktitle={International Workshop on Artificial Intelligence and Statistics},
  pages={333--340},
  year={2005},
  organization={PMLR}
}

@article{mertens2024retrieving,
  title={Retrieving the 21-cm signal from the Epoch of Reionization with learnt Gaussian process kernels},
  author={Mertens, Florent G and Bobin, J{\'e}r{\^o}me and Carucci, Isabella P},
  journal={MNRAS},
  volume={527},
  number={2},
  pages={3517--3531},
  year={2024},
  publisher={Oxford University Press}
}

@article{acharya202421,
  title={21-cm signal from the Epoch of Reionization: a machine learning upgrade to foreground removal with Gaussian process regression},
  author={Acharya, Anshuman and Mertens, Florent and Ciardi, Benedetta and Ghara, Raghunath and Koopmans, L{\'e}on VE and Giri, Sambit K and Hothi, Ian and Ma, Qing-Bo and Mellema, Garrelt and Munshi, Satyapan},
  journal={MNRAS},
  volume={527},
  number={3},
  pages={7835--7846},
  year={2024},
  publisher={Oxford University Press}
}

@article{munshi2025improved,
  title={Improved upper limits on the 21-cm signal power spectrum at z= 17.0 and z= 20.3 from an optimal field observed with NenuFAR},
  author={Munshi, S and Mertens, FG and Chege, JK and Koopmans, LVE and Offringa, AR and Semelin, B and Barkana, R and Dhandha, J and Fialkov, A and M{\'e}riot, R and others},
  journal={MNRAS},
  volume={542},
  number={4},
  pages={2785--2807},
  year={2025},
  publisher={Oxford University Press}
}

@article{vedantham2015scintillation,
  title={Scintillation noise in widefield radio interferometry},
  author={Vedantham, Harish K and Koopmans, LVE},
  journal={MNRAS},
  volume={453},
  number={1},
  pages={925--938},
  year={2015},
  publisher={Oxford University Press}
}

@article{vedantham2016scintillation,
  title={Scintillation noise power spectrum and its impact on high-redshift 21-cm observations},
  author={Vedantham, HK and Koopmans, LVE},
  journal={MNRAS},
  volume={458},
  number={3},
  pages={3099--3117},
  year={2016},
  publisher={Oxford University Press}
}

@article{brackenhoff2024ionospheric,
  title={Ionospheric contributions to the excess power in high-redshift 21-cm power-spectrum observations with LOFAR},
  author={Brackenhoff, SA and Mevius, M and Koopmans, LVE and Offringa, A and Ceccotti, E and Chege, JK and Gehlot, BK and Ghosh, S and H{\"o}fer, C and Mertens, FG and others},
  journal={MNRAS},
  volume={533},
  number={1},
  pages={632--656},
  year={2024},
  publisher={Oxford University Press}
}

@article{barry2016calibration,
  title={Calibration requirements for detecting the 21 cm epoch of reionization power spectrum and implications for the SKA},
  author={Barry, N and Hazelton, B and Sullivan, I and Morales, MF and Pober, JC},
  journal={MNRAS},
  volume={461},
  number={3},
  pages={3135--3144},
  year={2016},
  publisher={Oxford University Press}
}

@article{hofer2025impact,
  title={The impact of diffuse Galactic emission on direction-independent gain calibration in high-redshift 21 cm observations},
  author={H{\"o}fer, C and Koopmans, LVE and Brackenhoff, SA and Ceccotti, E and Chege, K and Ghosh, S and Mertens, FG and Mevius, M and Munshi, S and Offringa, AR},
  journal={arXiv preprint arXiv:2504.03554},
  year={2025}
}

@article{ceccotti2025spectral,
  title={Spectral modelling of Cygnus A between 110 and 250 MHz-Impact on the LOFAR 21-cm signal power spectrum},
  author={Ceccotti, E and Offringa, AR and Koopmans, LVE and Mertens, FG and Mevius, M and Acharya, A and Brackenhoff, SA and Ciardi, B and Gehlot, BK and Ghara, R and others},
  journal={A\&A},
  volume={696},
  pages={A56},
  year={2025},
  publisher={EDP Sciences}
}

@article{gehlot2024transient,
  title={Transient RFI environment of LOFAR-LBA at 72--75 MHz-Impact on ultra-widefield AARTFAAC Cosmic Explorer observations of the redshifted 21-cm signal},
  author={Gehlot, BK and Koopmans, LVE and Brackenhoff, SA and Ceccotti, E and Ghosh, S and H{\"o}fer, C and Mertens, FG and Mevius, M and Munshi, S and Offringa, AR and others},
  journal={A\&A},
  volume={681},
  pages={A71},
  year={2024},
  publisher={EDP Sciences}
}

@article{munshi2025near,
  title={Near-field imaging of local interference in radio interferometric data-Impact on the redshifted 21 cm power spectrum},
  author={Munshi, S and Mertens, FG and Koopmans, LVE and Mevius, M and Offringa, AR and Semelin, B and Viou, C and Bracco, A and Brackenhoff, SA and Ceccotti, E and others},
  journal={A\&A},
  volume={697},
  pages={A203},
  year={2025},
  publisher={EDP Sciences}
}

@article{wilensky2019absolving,
  title={Absolving the SSINS of precision interferometric radio data: a new technique for mitigating faint radio frequency interference},
  author={Wilensky, Michael J and Morales, Miguel F and Hazelton, Bryna J and Barry, Nichole and Byrne, Ruby and Roy, Sumit},
  journal={Publications of the Astronomical Society of the Pacific},
  volume={131},
  number={1005},
  pages={114507},
  year={2019},
  publisher={IOP Publishing}
}

@article{offringa2019impact,
  title={The impact of interference excision on 21-cm epoch of reionization power spectrum analyses},
  author={Offringa, AR and Mertens, F and Koopmans, LVE},
  journal={MNRAS},
  volume={484},
  number={2},
  pages={2866--2875},
  year={2019},
  publisher={Oxford University Press}
}

@article{munshi2025beyond,
  title={Beyond the horizon: Quantifying the full sky foreground wedge in the cylindrical power spectrum},
  author={Munshi, S and Mertens, FG and Koopmans, LVE and Offringa, AR and Ceccotti, E and Brackenhoff, SA and Chege, JK and Gehlot, BK and Ghosh, S and H{\"o}fer, C and others},
  journal={A\&A},
  volume={693},
  pages={A276},
  year={2025},
  publisher={EDP Sciences}
}

@article{vedantham2012imaging,
  title={Imaging the Epoch of Reionization: limitations from foreground confusion and imaging algorithms},
  author={Vedantham, Harish and Shankar, N Udaya and Subrahmanyan, Ravi},
  journal={ApJ},
  volume={745},
  number={2},
  pages={176},
  year={2012},
  publisher={IOP Publishing}
}

@article{morales2012four,
  title={Four fundamental foreground power spectrum shapes for 21 cm cosmology observations},
  author={Morales, Miguel F and Hazelton, Bryna and Sullivan, Ian and Beardsley, Adam},
  journal={ApJ},
  volume={752},
  number={2},
  pages={137},
  year={2012},
  publisher={IOP Publishing}
}

@article{datta2010bright,
  title={Bright source subtraction requirements for redshifted 21 cm measurements},
  author={Datta, A and Bowman, JD and Carilli, CL},
  journal={ApJ},
  volume={724},
  number={1},
  pages={526},
  year={2010},
  publisher={IOP Publishing}
}

@article{gan2022statistical,
  title={Statistical analysis of the causes of excess variance in the 21 cm signal power spectra obtained with the Low-Frequency Array},
  author={Gan, H and Koopmans, LVE and Mertens, FG and Mevius, M and Offringa, AR and Ciardi, B and Gehlot, BK and Ghara, R and Ghosh, A and Giri, SK and others},
  journal={A\&A},
  volume={663},
  pages={A9},
  year={2022},
  publisher={EDP Sciences}
}

@article{gan2023assessing,
  title={Assessing the impact of two independent direction-dependent calibration algorithms on the LOFAR 21 cm signal power spectrum-And applications to an observation of a field flanking the north celestial pole},
  author={Gan, H and Mertens, FG and Koopmans, LVE and Offringa, AR and Mevius, M and Pandey, VN and Brackenhoff, Stefanie A and Ceccotti, E and Ciardi, B and Gehlot, BK and others},
  journal={A\&A},
  volume={669},
  pages={A20},
  year={2023},
  publisher={EDP Sciences}
}

@article{di2023unintended,
  title={Unintended electromagnetic radiation from Starlink satellites detected with LOFAR between 110 and 188 MHz},
  author={Di Vruno, F and Winkel, B and Bassa, CG and J{\'o}zsa, GIG and Brentjens, MA and Jessner, A and Garrington, S},
  journal={A\&A},
  volume={676},
  pages={A75},
  year={2023},
  publisher={EDP Sciences}
}

@article{grigg2023detection,
  title={Detection of intended and unintended emissions from Starlink satellites in the SKA-Low frequency range, at the SKA-Low site, with an SKA-Low station analogue},
  author={Grigg, Dylan and Tingay, SJ and Sokolowski, Marcin and Wayth, RB and Indermuehle, Balthasar and Prabu, Steve},
  journal={A\&A},
  volume={678},
  pages={L6},
  year={2023},
  publisher={EDP Sciences}
}

@article{offringa2019precision,
  title={Precision requirements for interferometric gridding in the analysis of a 21 cm power spectrum},
  author={Offringa, AR and Mertens, F and Van der Tol, S and Veenboer, B and Gehlot, BK and Koopmans, LVE and Mevius, M},
  journal={A\&A},
  volume={631},
  pages={A12},
  year={2019},
  publisher={EDP Sciences}
}

@phdthesis{matern1960spatial,
  title={Spatial variation: Stachastic models and their application to some problems in forest surveys and other sampling investigations},
  author={Mat{\'e}rn, Bertil},
  year={1960},
  school={Stockholm University}
}

@book{journel1976mining,
  title={Mining geostatistics},
  author={Journel, Andre G and Huijbregts, Charles J},
  year={1976}
}

@article{ghosh2020foreground,
  title={Foreground modelling via Gaussian process regression: an application to HERA data},
  author={Ghosh, Abhik and Mertens, Florent and Bernardi, Gianni and Santos, M{\'a}rio G and Kern, Nicholas S and Carilli, Christopher L and Grobler, Trienko L and Koopmans, L{\'e}on VE and Jacobs, Daniel C and Liu, Adrian and others},
  journal={MNRAS},
  volume={495},
  number={3},
  pages={2813--2826},
  year={2020},
  publisher={Oxford University Press}
}

@article{parsons2012per,
  title={A per-baseline, delay-spectrum technique for accessing the 21 cm cosmic reionization signature},
  author={Parsons, Aaron R and Pober, Jonathan C and Aguirre, James E and Carilli, Christopher L and Jacobs, Daniel C and Moore, David F},
  journal={ApJ},
  volume={756},
  number={2},
  pages={165},
  year={2012},
  publisher={IOP Publishing}
}

@article{ceccotti2025first,
    author = {Ceccotti, E and Offringa, A R and Mertens, F G and Koopmans, L V E and Munshi, S and Chege, J K and Acharya, A and Brackenhoff, S A and Chapman, E and Ciardi, B and Ghara, R and Ghosh, S and Giri, S K and Höfer, C and Hothi, I and Mellema, G and Mevius, M and Pandey, V N and Zaroubi, S},
    title = {First upper limits on the 21-cm signal power spectrum of neutral hydrogen at z = 9.16 from the LOFAR 3C 196 field},
    journal = {MNRAS},
    pages = {staf1629},
    year = {2025},
    month = {09},
    issn = {0035-8711},
    doi = {10.1093/mnras/staf1629},
    url = {https://doi.org/10.1093/mnras/staf1629},
    eprint = {https://academic.oup.com/mnras/advance-article-pdf/doi/10.1093/mnras/staf1629/64415860/staf1629.pdf},
}

@article{acharya2024revised,
  title={Revised LOFAR upper limits on the 21-cm signal power spectrum at z≈ 9.1 using machine learning and gaussian process regression},
  author={Acharya, Anshuman and Mertens, Florent and Ciardi, Benedetta and Ghara, Raghunath and Koopmans, L{\'e}on VE and Zaroubi, Saleem},
  journal={MNRAS: Letters},
  volume={534},
  number={1},
  pages={L30--L34},
  year={2024},
  publisher={Oxford University Press}
}

@article{dewdney2009square,
  title={The square kilometre array},
  author={Dewdney, Peter E and Hall, Peter J and Schilizzi, Richard T and Lazio, T Joseph LW},
  journal={Proceedings of the IEEE},
  volume={97},
  number={8},
  pages={1482--1496},
  year={2009},
  publisher={IEEE}
}

@article{10.1093/mnras/staf1466,
    author = {Bonaldi, A and Hartley, P and Braun, R and Purser, S and Acharya, A and Ahn, K and Aparicio Resco, M and Bait, O and Bianco, M and Chakraborty, A and Chapman, E and Chatterjee, S and Chege, K and Chen, H and Chen, X and Chen, Z and Conaboy, L and Cruz, M and Darriba, L and De Santis, M and Denzel, P and Diao, K and Feron, J and Finlay, C and Gehlot, B and Ghosh, S and Giri, S K and Grumitt, R and Hong, S E and Ito, T and Jiang, M and Jordan, C and Kim, S and Kim, M and Kim, J and Krishna, S P and Kulkarni, A and López-Caniego, M and Labadie-García, I and Lee, H and Lee, D and Lee, N and Line, J and Liu, Y and Mao, Y and Mazumder, A and Mertens, F G and Munshi, S and Nasirudin, A and Ni, S and Nistane, V and Norregaard, C and Null, D and Offringa, A and Oh, M and Oh, S-H and Parkinson, D and Pritchard, J and Ruiz-Granda, M and Salvador López, V and Shan, H and Sharma, R and Trott, C and Yoshiura, S and Zhang, L and Zhang, X and Zheng, Q and Zhu, Z and Zuo, S and Akahori, T and Alberto, P and Allys, E and An, T and Anstey, D and Baek, J and Basavraj and Brackenhoff, S and Browne, P and Ceccotti, E and Chen, H and Chen, T and Choudhuri, S and Choudhury, M and Coles, J and Cook, J and Cornu, D and Cunnington, S and Das, S and de Lera Acedo Acedo, E and Delouis, J-M and Deng, F and Ding, J and Elahi, K M A and Fernandez, P and Fernández, C and Fernández Alcázar, A and Galluzzi, V and Gao, L-Y and Garain, U and Garrido, J and Gendron-Marsolais, M-L and Gessey-Jones, T and Ghorbel, H and Gong, Y and Guo, S and Hasegawa, K and Hayashi, T and Herranz, D and Holanda, V and Holloway, A J and Hothi, I and Höfer, C and Jelić, V and Jiang, Y and Jiang, X and Kang, H and Kim, J-Y and Koopmans, L V and Lacroix, R and Lee, E and Leeney, S and Levrier, F and Li, Y and Liu, Y and Ma, Q and Meriot, R and Mesinger, A and Mevius, M and Minoda, T and Miville-Deschênes, M-A and Moldon, J and Mondal, R and Murmu, C and Murray, S and Nirmala SR and Niu, Q and Nunhokee, C and O’Hara, O and Pal, S K and Pal, S and Park, J and Parra, M and Patra, N N and Pindor, B and Remazeilles, M and Rey, P and Rubino-Martin, J A and Saha, S and Selvaraj, A and Semelin, B and Shah, R and Shao, Y and Shaw, A K and Shi, F and Shimabukuro, H and Singh, G and Sohn, B W and Stagni, M and Starck, J-L and Sui, C and Swinbank, J D and Sánchez, J and Sánchez-Expósito, S and Takahashi, K and Takeuchi, T and Tripathi, A and Verdes-Montenegro, L and Vielva, P and Vitello, F R and Wang, G-J and Wang, Q and Wang, X and Wang, Y and Wang, Y-X and Wiegert, T and Wild, A and Williams, W L and Wolz, L and Wu, X and Wu, P and Xia, J-Q and Xu, Y and Yan, R and Yan, Y-P and Yin, Z and You, Z and Yu, X and Yu, K and Yue, B and Zhang, L and Zhao, Z and Zhao, X and Zhou, X},
    title = {Square Kilometre Array Science Data Challenge 3a: foreground removal for an EoR experiment},
    journal = {MNRAS},
    volume = {543},
    number = {2},
    pages = {1092-1119},
    year = {2025},
    month = {09},
    issn = {0035-8711},
    doi = {10.1093/mnras/staf1466},
    url = {https://doi.org/10.1093/mnras/staf1466},
    eprint = {https://academic.oup.com/mnras/article-pdf/543/2/1092/64222444/staf1466.pdf},
}

\appendix
\section{The linear model of co-regionalisation}\label{sec:lmc}
In this section, we first describe the LMC framework outlined by \cite{alvarez2012kernels}, and then show how the cross-GPR formalism developed in Sect.~\ref{sec:joint_gpr_multiple} connects to it. We consider a multi-output GP with $D$ outputs, $Q$ groups of latent functions $\{u_q(\mathbf{x})\}_{q=1}^{Q}$ each corresponding to a different covariance $\mathbf{K}_q$, with the $q-$th group consisting of $R_q$ functions $\{u_q^i(\mathbf{x})\}_{i=1}^{R_q}$ which share the same covariance, but are independent. In the LMC framework, the functions and the cross-covariance between the $d-$th and the $d'-$th outputs are given by
\begin{align}\label{eq:lmc}
&f_d(\mathbf{x})=\sum_{q=1}^Q \sum_{i=1}^{R_q} a_{d, q}^i u_q^i(\mathbf{x}),\nonumber\\
&\mathbf{K}^{d,d'} = \sum_{q=1}^{Q} b^q_{d,d'} \mathbf{K}_q, \text{ where } b_{d, d^{\prime}}^q=\sum_{i=1}^{R_q} a_{d, q}^i a_{d^{\prime}, q}^i .
\end{align}
The coefficients $b^q_{d,d'}$ can be written as elements of the matrix $\mathbf{B}_q \in \mathbb{R}^{D\times D}$, called a co-regionalisation matrix. The full covariance $\mathbf{K}$ for the $D$ outputs can then be written in the form
\begin{equation}
\mathbf{K} = \sum_{q=1}^Q \mathbf{B}_q \otimes \mathbf{K}_q.
\end{equation}
In our case, given by Eq.~(\ref{eq:joint_funcs}), the data consists of a coherent and an incoherent component, with two datasets. We let $q=1$ correspond to the coherent component and $q=2$ correspond to the incoherent component. The cross-GPR formalism for two datasets now can be written as a specific case of the LMC framework, with $D=2$, $Q=2$, $R_1=1$, and $R_2=2$. Comparing Eq.~(\ref{eq:joint_funcs}) to Eq.~(\ref{eq:lmc}), the functions, covariances, and coefficients are then given by
\begin{align}
&u_1(\mathbf{x}) = \mathbf{f}_{\mathrm{coh}},\ u_2^1(\mathbf{x}) = \mathbf{f}_{\mathrm{inc}}^1,\ u_2^2(\mathbf{x}) = \mathbf{f}_{\mathrm{inc}}^2,\nonumber\\
&\mathbf{K}_1 = \mathbf{K}_\mathrm{coh}, \ \mathbf{K}_2 = \mathbf{K}_\mathrm{inc}, \nonumber\\
&a_{1,1}=a_{2,1}=1, a_{1,2}^1=a_{2,2}^2=1, a_{1,2}^2=a_{2,2}^1=0.
\end{align}
Inserting these coefficients into Eq.~(\ref{eq:lmc}), for pairs of $d,d'$, the co-regionalisation matrices, $\mathbf{B}_q$, described by the coefficients $b^q_{d,d'}$, are obtained
\begin{align}
&B_1 = \begin{bmatrix}
b^1_{1,1} & b^1_{1,2}\\
b^1_{2,1} & b^1_{2,2}
\end{bmatrix}
= \begin{bmatrix}
1 & 1\\
1 & 1
\end{bmatrix},\nonumber\\
&B_2 = \begin{bmatrix}
b^2_{1,1} & b^2_{1,2}\\
b^2_{2,1} & b^2_{2,2}
\end{bmatrix}
= \begin{bmatrix}
1 & 0\\
0 & 1
\end{bmatrix}.
\end{align}
The model covariance matrix is then given by
\begin{align}
\mathbf{K}=\begin{bmatrix}
    \mathbf{K}_\mathrm{coh}+\mathbf{K}_\mathrm{inc} & \mathbf{K}_\mathrm{coh}\\
    \mathbf{K}_\mathrm{coh} & \mathbf{K}_\mathrm{coh}+\mathbf{K}_\mathrm{inc}
\end{bmatrix}.
\end{align}
This retrieves the covariance model structure obtained in Eq.~(\ref{eq:joint_cross}).

\end{document}